\documentstyle[11pt]{article}
\newfont{\larom}{cmbx10 scaled\magstep3}
\newfont{\lar}{cmbx10 scaled\magstephalf}
\newfont{\bsan}{cmssbx10}
\begin{document}
\input{epsf.tex}
\setcounter{page}{1}  
\begin{center}
  {\larom FRACTALS AND THE DISTRIBUTION OF GALAXIES} 

  \vspace{10mm}

  {Marcelo B.\ Ribeiro\footnote{ \ Present address: \it Dept.\ F\'{\i}sica
  Matem\'atica, Instituto de F\'{\i}sica--UFRJ, Caixa Postal 68528, Ilha do
  Fund\~ao, Rio de Janeiro, RJ 21945-970, Brazil; E-mail: mbr@if.ufrj.br.}
  \& Alexandre Y. Miguelote\footnote{ \ Former participant of the PIBIC
  program at Observat\'{o}rio Nacional--CNPq. Present address: \it
  Engenharia de Sistemas e Computa\c c\~ao, Coppe-UFRJ, Ilha do Fund\~ao,
  Rio de Janeiro, Brazil; E-mail: yasuda@cos.ufrj.br.}\\~\\ \it
  Observat\'{o}rio Nacional--CNPq, Rio de Janeiro, Brazil} 

%

  \vspace{5mm}
\end{center}
\begin{quotation}

{\bf Abstract:} This paper presents a review of the fractal approach for
describing the large scale distribution of galaxies. We start by presenting
a brief, but general, introduction to fractals, which emphasizes their
empirical side and applications rather than their mathematical side. Then
we discuss the standard correlation function analysis of galaxy catalogues
and many observational facts that brought in\-creas\-ing doubts about the
reliability of this method, paying special attention to the standard analysis
implicit assumption of an eventual homogeneity of the distribution of
galaxies. Some new statistical concepts for analysing this distribution is
presented, and without the implicit assumption of homogeneity they bring
support to the hypothesis that the distribution of galaxies does form a
fractal system. The Pietronero-Wertz's single fractal (hierarchical) model
is presented and discussed, together with the implications of this new
approach for understanding galaxy clustering.

\vspace{3mm}

\end{quotation}
\section{\bf Introduction}\label{Introduction}

The goal of modern cosmology is to find the large scale matter
distribution and spacetime structure of the Universe from astronomical
observations, and it dates back from the early days of observational
cosmology the realization that in order to achieve this aim it is
essential that an accurate empirical description of galaxy clustering
be derived from the systematic observations of distant galaxies. As time
has passed, this realization has become a program, which in the last decade
or so took a great impulse forward due to improvements in astronomical
data acquisition techniques and data analysis. As a result of that
an enormous amount of data about the observable universe was accumulated
in the form of the now well-known {\it redshift surveys}, and some
widely accepted conclusions drawn from these data created a certain
confidence in many researchers that such an accurate description of the
distribution of galaxies was just about to being achieved. However, those
conclusions are mainly based in a standard statistical analysis derived
from a scenario provided by the standard Friedmannian cosmological models,
which assume homogeneity and isotropy of the matter distribution, scenario
which is still thought by many to be the best theoretical framework capable
of explaining the large scale matter distribution and spacetime structure
of the Universe. 

The view outlined above, which now has become the orthodox homogeneous
universe view, has, however, never been able to fully overcome some of
its objections. In particular, many researchers felt in the past, and
others still feel today, that the relativistic derived idea of an
eventual homogenization of the {\it observed} matter distribution of the
Universe is flawed, since, in their view, the empirical evidence collected
from the systematic observation of distant cosmological sources also
supports the claim that the universal distribution of matter will not
eventually homogenize. Therefore, the critical voice claims that the
large-scale distribution of matter in the Universe is intrinsically 
inhomogeneously distributed, from the smallest to the largest observed
scales and, perhaps, indefinitively.

Despite this, it is a historical fact that the inhomogeneous view has
never been as developed as the orthodox view, and perhaps the major cause 
for this situation was the lack of workable models supporting this
inhomogeneous claim. There has been, however, one major exception, in the
form of a hierarchical cosmological model advanced by Wertz (1970, 1971),
although, for reasons that will be explained below, it has unfortunately
remained largely ignored so far. 

Nevertheless, by the mid 1980's those objections took a new vigour
with the arrival of a new method for describing galaxy clustering
based on ideas of a radically new geometrical perspective for
the description of irregular patterns in nature: {\it the fractal
geometry}.

In this review we intend to show the basic ideas behind this new
approach for the galaxy clustering problem. We will not present the
orthodox traditional view since it can be easily found, for instance,
in Peebles (1980, 1993) and Davis (1997). Therefore, we shall
concentrate ourselves in the challenging voice based on a new
viewpoint about the statistical characterization of galaxy clustering,
whose results go against many traditional expectations, and which keep
open the possibility that the universe never becomes observationally
homogeneous. The basic papers where this fractal view for the
distribution of galaxies can be found are relatively recent. Most of
what will be presented here is based on Pietronero (1987), Pietronero,
Montuori and Sylos Labini (1997), and on the comprehensive reviews by
Coleman and Pietronero (1992), and Sylos Labini, Montuori and
Pietronero (1998).

The plan of the paper is as follows. In section 2 we present a brief, but
general, introduction to fractals, which emphasizes their empirical
side and applications, but without neglecting their basic mathematical
concepts. Section \ref{Distr Galaxies} briefly presents the basic current
analysis of the large scale distribution of galaxies, its difficulties
and, finally, Pietronero-Wertz's single fractal (hierarchical) model that
proposes an alternative point of view for describing and analysing this
distribution, as well as some of the consequences of such an approach.
The paper finishes with a discussion on some aspects of the current
controversy about the fractal approach for describing the distribution of
galaxies. 

\section{Fractals}\label{Fractals}

This section introduces a minimum background material on fractals
necessary in this paper and which may be useful for readers not
familiar with their basic ideas and methods. Therefore, we shall not
present an extensive, let alone comprehensive, discussion of the
subject, which can be found in Mandelbrot (1983), Feder (1988),
Takayasu (1990) and Peitgen, J\"{u}rgens and Saupe (1992), if the
reader is more interested in the intuitive
notions associated with fractals and their applications, or in Barnsley
(1988) and Falconer (1990) if the interest is more mathematical. The
literature on this subject is currently growing at a bewildering pace
and those books represent just a small selection that can be used for
different purposes when dealing with fractals. This section consists
mainly of a summary of a background material basically 
selected from these sources. The discussion starts on the mathematical 
aspects associated with fractals, but gradually there is a growing 
emphasis on applications.

\subsection{On the ``Definition'' of Fractals}\label{Def of Fracalts}

The name {\it fractal} was introduced by Benoit B. Mandelbrot to
characterize geometrical figures which are not smooth or regular. By
adopting the saying in Latin that {\it nomen est numen}, \footnote{ \ To
name is to know.} he decided to ``exert the right of naming newly
opened or newly settled territory landmarks.'' Thus, he ``coined
{\it fractal} from the Latin adjective {\it fractus}. The corresponding
Latin verb {\it frangere} means `to break:' to create irregular
fragments. It is therefore sensible (...) that, in addition to
`fragmented' (as in {\it fraction} or
{\it refraction}), {\it fractus} should also mean `irregular', both
meanings being preserved in {\it fragment} '' (Mandelbrot 1983, p. 4).

In attempting to define fractals mathematically, Mandelbrot (1983)
offered the following: ``A fractal is by definition a set for which
the Hausdorff-Besicovitch dimension strictly exceeds the topological
dimension''. We shall discuss later the Hausdorff dimension, but the
important point here is that Mandelbrot himself has since retreated from
this original tentative definition as it proved to be unsatisfactory, in
the sense that it excluded some sets which ought to be regarded as
fractals.  In a private communication to J. Feder, Mandelbrot proposed
instead the following loose tentative definition: ``a fractal is a shape 
made of parts similar to the whole in some way'' (Feder 1988, p. 11).
Even so, in the school on ``Fractal Geometry and Analysis'' that took
place in Montreal in 1989, and 
was attended by one of us (MBR) \footnote{ \ See B\'elair 
and Dubuc (1991).}, Mandelbrot declined to discuss the problem of definition 
by arguing that any one would be restrictive and, perhaps, it would be best to 
consider fractals as a collection of techniques and methods applicable 
in the study of these irregular, broken and self-similar geometrical patterns.

Falconer (1990) offers the following point of view on this matter.
\begin{quotation}
 \small

``The definition of a `fractal' should be regarded in the same way as the
biologist regards the definition of `life'. There is no hard and fast
definition, but just a list of properties characteristic of a living
thing, such as the ability to reproduce or to move or to exist to some
extent independently of the environment. Most living things have most of
the characteristics on the list, though there are living objects that
are exceptions to each of them. In the same way, it seems best to regard
a fractal as a set that has properties such as those listed below,
rather than to look for a precise definition which will almost
certainly exclude some interesting cases. (...) When we refer to a set
$F$ as a fractal, therefore, we will typically have the following in
mind.

\begin{enumerate}
\item $F$ has a fine structure, i.e., detail on arbitrarily small scales.
\item $F$ is too irregular to be described in traditional geometrical
      language, both locally and globally.
\item Often $F$ has some form of self-similarity, perhaps approximate or
      statistical.
\item Usually, the `fractal dimension' of $F$ (defined in some way) is
      greater than its topological dimension.
\item In most cases of interest $F$ is defined in a very simple way,
      perhaps recursively.''
\end{enumerate}

\end{quotation}

It is clear that Falconer's ``definition'' of fractals includes
Mandelbrot's loose tentative one quoted above.
Moreover, Falconer keeps open the possibility that a given fractal shape
can be characterized by more than one definition of fractal dimension,
and they do not necessarily need to coincide with each other, although
they have in common the property of being able to take fractional
values. Therefore, an important aspect of studying a fractal structure
(once it is characterized as such by, say, at least being recognized as
self-similar in some way) is the choice of a definition for fractal
dimension that best applies to, or is derived from, the case in study.
As the current trend appears to indicate that this absence of a strict
definition for fractals will prevail, the word fractal can be, and in
fact is, even among specialists, used as a generic noun, and sometimes
as an adjective.

Fractal geometry has been considered a revolution in the way we are able
to mathematically represent and study figures, sets and functions. In
the past sets or functions that are not sufficiently smooth or regular
tended to be ignored as ``pathological'' or considered as mathematical
``monsters'',
and not worthy of study, being regarded only as individual curiosities.
Nowadays, there is a realization that a lot can and is worth being said
about non-smooth sets. Besides, irregular and broken sets provide a much
better representation of many phenomena than do the figures of classical
geometry.

\subsection{The Hausdorff Dimension}\label{Haus Dim}

An important step in the understanding of fractal dimensions is for one
to be introduced to the {\it Hausdorff-Besicovitch dimension} (often known
simply as {\it Hausdorff dimension}). It was first introduced by F. Hausdorff
in 1919 and developed later in the 1930's by A. S. Besicovitch and his
students. It can take non-integer values and was found to coincide with
many other definitions. 

In obtaining the dimension that bears his name, Hausdorff used the idea
of defining measures using covers of sets first proposed by C.\
Carath\'{e}odory in 1914. Here in this article, however, we shall avoid a
formal mathematical demonstration of the Hausdorff measure, and keep the
discussion in intuitive terms. We shall therefore offer an {\it
illustration} of the Hausdorff measure whose final result is the same as
achieved by the formal proof found, for instance, in Falconer (1990).

\begin{figure}[thb]
  \centerline{\epsffile{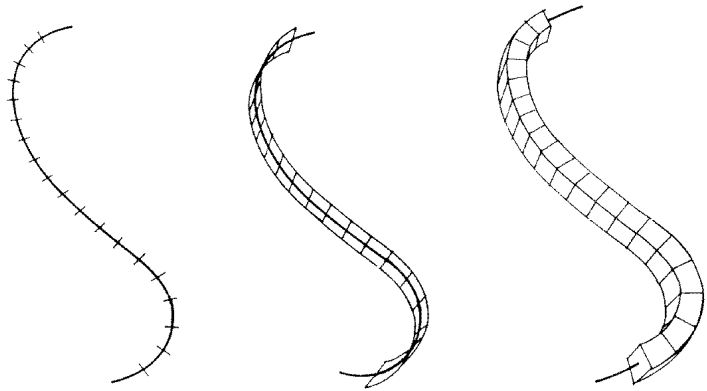}}
  \caption{\sf Measuring the ``size'' of curves (Feder 1988).}\label{2-1}
\end{figure}

The basic question to answer is: how do we measure the ``size'' of a set
$F$ of points in space? A simple manner of measuring the length of
curves, the area of surfaces or the volume of objects is to divide the 
space into small cubes of diameter $\delta$ as shown in figure \ref{2-1}.
Small spheres of diameter $\delta$ could have been used instead. Then the
curve can be measured by finding the number $N(\delta)$ of line segments
of length $\delta$ needed to cover the line. Obviously for an
ordinary curve we have $N(\delta)=L_\ast / \delta$. The length of the
curve is given by
\[
  L = N(\delta) \delta \mathop\rightarrow_{\delta \to 0} 
      L_\ast \delta^0. 
\]
In the limit $\delta \rightarrow 0$, the measure $L$ becomes
asymptotically equal to the length of the curve and is independent of
$\delta$.

In a similar way we can associate an area with the set of points
defining a curve by obtaining the number of disks or squares needed to
cover the curve. In the case of squares where each one has an area of
$\delta^2$, the number of squares $N(\delta)$ gives an associated area 
\[
  A = N(\delta) \delta^2 \mathop\rightarrow_{\delta \to 0}
      L_\ast \delta^1.
\]
In a similar fashion the volume $V$ associated with the line is given by
\[
  V = N(\delta) \delta^3  \mathop\rightarrow_{\delta \to 0}
      L_\ast \delta^2.
\]
Now, for ordinary curves both $A$ and $V$ tend to zero as $\delta$
vanishes, and the only interesting measure is the length of the curve.

\begin{figure}[thb]
  \centerline{\epsffile{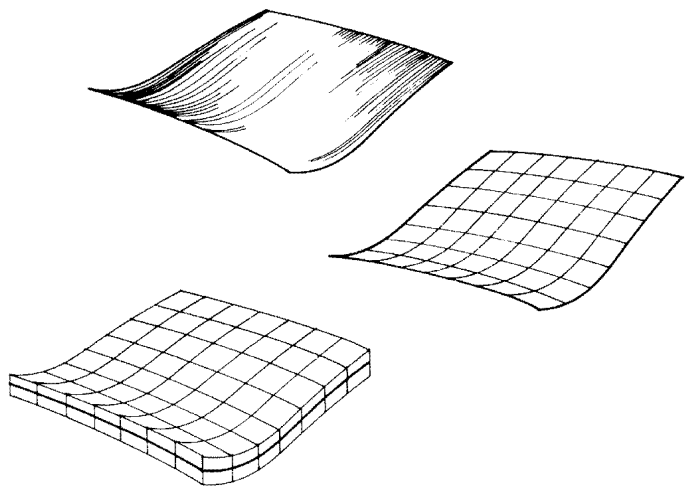}}
  \caption{\sf Measuring the ``size'' of a surface (Feder 1988).}\label{2-2}
\end{figure}

Let us consider now a set of points that define a surface as illustrated
in figure \ref{2-2}. The normal measure is the area $A$, and so we have
\[
  A = N(\delta) \delta^2  \mathop\rightarrow_{\delta \to 0}
      A_\ast \delta^0.
\]
Here we find that for an ordinary surface the number of squares needed
to cover it is $N(\delta) = A_\ast / \delta^2$ in limit of vanishing
$\delta$, where $A_\ast$ is the area of the surface. We may associate a
volume with the surface by forming the sum of the volumes of the cubes
needed to cover the surface:
\[
  V = N(\delta) \delta^3 \mathop\rightarrow_{\delta \to 0}
      A_\ast \delta^1.
\]
This volume vanishes as $\delta \rightarrow 0$, as expected.

Now, can we associate a {\it length} with a surface? Formally we can
simply take the measure
\[
  L = N(\delta) \delta \rightarrow  A_\ast \delta^{-1},
\]
which diverges for $\delta \rightarrow 0$. This is a reasonable result
since it is impossible to cover a surface with a finite number of line
segments. We conclude that the only useful measure of a set of points
defined by a surface in three-dimensional space is the area.

There are curves, however, that twist so badly that their length is
infinite, but they are such that they fill the plane (they have the
generic name of {\it Peano curves}~\footnote{ \ See \S \protect\ref{Sim 
Dim Peano Curves}}). Also there are surfaces that fold so wildly that 
they fill the space. We can discuss such strange sets of points if we 
generalize the measure of size just discussed.

\begin{figure}[thb]
  \centerline{\epsffile{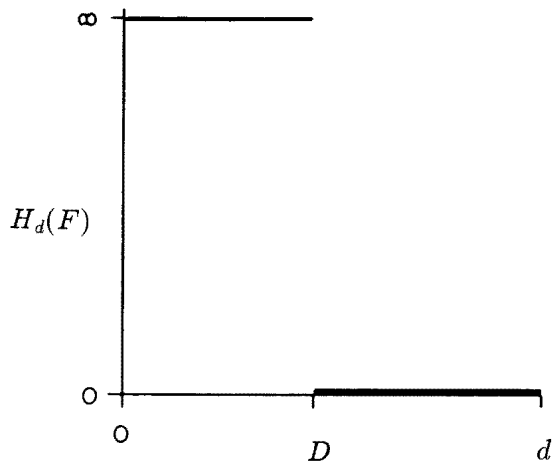}}
  \caption[\sf Graph of the Hausdorff $d$-measure $H_d(F)$ against $d$ for a
           set $F$ (Falconer 1990).]{\sf Graph of $H_d(F)$ against $d$
	   for a set $F$. The Hausdorff dimension is the critical value
	   of $d$ at which the jump from $\infty$ to $0$ occurs
	   (Falconer 1990).}\label{2-3}
\end{figure}

So far, in order to give a measure of the size of a set $F$ of points, 
in space we have taken a test function $h(\delta) = \gamma (d) \delta^d$
-- a line, a square, a disk, a ball or a cube -- and have covered the
set to form the {\it measure} $H_d(F) = \sum h(\delta)$. For lines,
squares and cubes we have the geometrical factor $\gamma (d) = 1$. In
general we find that, as $\delta \rightarrow 0$, the measure $H_d(F)$ is
either infinite or zero depending on the choice of $d$, the dimension of
the measure. The Hausdorff dimension $D$ of the set $F$ is the {\it
critical dimension} for which the measure $H_d(F)$ jumps from infinity
to zero (see figure \ref{2-3}):
\begin{equation}
  H_d(F) = \sum \gamma (d) \delta^d = \gamma (d) N(\delta) \delta^d
	   \mathop\rightarrow_{\delta \to 0}
	   \left\{ \begin{array}{ll}
                      0,      &  d > D, \\
                      \infty, &  d < D.
                   \end{array}
           \right.
  \label{eq-ch2-1}
\end{equation}
The quantity $H_d(F)$ is called the $d$-measure of the set and its value
for $d=D$ is often finite, but may be zero or infinity. It is the
position of the jump in $H_d$ as function of $d$ that is important. Note
that this definition means the Hausdorff dimension $D$ is a local
property in the sense that it measures properties of sets of points in
the limit of a vanishing diameter of size $\delta$ of the test function
used to cover the set. It also follows that the $D$ may depend on
position. There are several points and details to be considered, but
they are not important for us here (see Falconer 1990).

The familiar cases are $D=1$ for lines, $D=2$ for planes and surfaces
and $D=3$ for spheres and other finite volumes. There are innumerable
sets, however, for which the Hausdorff dimension is noninteger and is
said to be fractal. In other words, because the jump of the measure
$H_d(F)$ can happen at noninteger values of $d$, when $H_d(F)$ is
calculated for irregular and broken sets the value $D$ where the jump
actually occurs is usually noninteger.

\subsection{Other Fractal Dimensions and Some Examples of \protect\newline
 Fractals}\label{Other Frac Dim}

The illustration of the Hausdorff dimension shown previously 
may be very well from a mathematical point of view, but it is hard
to get an intuition about the fractal dimension from it. Moreover, we do
not have a clear picture of what this fractional value of dimension
means. In order to try to answer these questions let us see  different
definitions of fractal dimension and some examples.

\subsubsection{Similarity Dimension and Peano Curves}\label{Sim Dim Peano
Curves}

The fractals we discuss may be considered to be sets of points embedded
in space. This space has the usual topological dimension which we are
used to, and from a physicist's point of view it coincides with
degrees of freedom defined by the number of independent variables. So
the location of a point on a line is determined by one real number and a
set of two independent real numbers is needed to define a plane. If we
define dimension by degrees of freedom in this way, we can consider a
$d$-dimensional space for any non-negative integer $d$. In fact, in
mechanics it is conventional to consider the motion of $m$ particles in
3 dimensions as being the motion of one particle in a $6m$-dimensional
space if we take each particle's location and momentum as independent.

The dimension defined by degrees of freedom seems very natural, but
more than 100 years ago it was found to contain a serious flaw. In 1890
Giuseppe Peano described a curve that folds so wildly that it ``fills'' the
plane. We shall describe below what we mean by {\it plane-filling curves}
and how we can generate them, but let us  keep this meaning in a
qualitative and intuitive form for the moment. The important point is
that those {\it Peano curves} can be drawn with a single stroke and they
tend to cover the plane uniformly, being able not only to avoid
self-intersection but also self-contact in some cases. Since the
location of a point
on the Peano curve, like a point in any curve, can be characterized with
one real number, we become able to describe the position of any point on
a plane with only one real number. Hence the degree of freedom, or the
dimension, of this plane becomes 1, which contradicts the empirical
value 2. This process of natural parameterization produced by the
Peano ``curves'' was called {\it Peano motions} or {\it plane-filling
motions} by Mandelbrot (1983, p. 58).

\begin{figure}[thb]
  \centerline{\epsffile{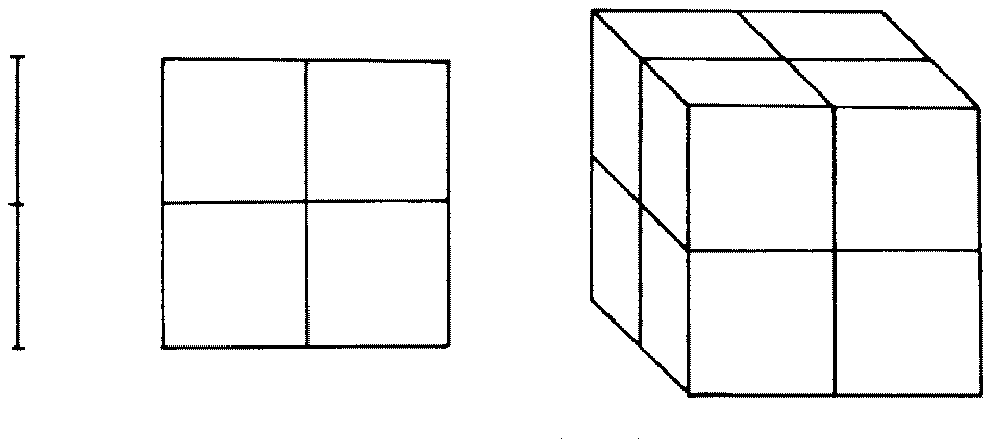}}
  \caption{\sf Dividing a segment, a square and a cube
	   (Takayasu 1990).}\label{2-4}
\end{figure}

In order to see how Peano curves fill the plane, let us introduce first
a new definition of dimension based on similarity. If we divide an unit
line segment into $N$ parts (see figure \ref{2-4}) we get at the end of
the process each part of the segment scaled by  a factor $\delta = 1/N$,
which means $N \delta^1 = 1$. If we divide an unit square into $N$
similar parts, each one is scaled by a factor $\delta = 1/N^{1/2}$ (if
$N=4$ the square is scaled by half the side length); so $N \delta^2 =1$.
Now if an unit cube is divided in $N$ parts, each scaled by $\delta =
1/N^{1/3}$ (again if $N=8$ the cube is scaled by half the side length),
we have $N \delta^3 =1$. Note that the exponents of $\delta$ correspond
to the space dimensions in each case. Generalising this discussion we
may say that for an object of $N$ parts, each scaled down from the whole
by a ratio $\delta$, the relation $N \delta^D =1$ defines the {\it
similarity dimension} $D$ of the set as
\begin{equation}
   D= \frac{ \log N}{ \log 1/\delta}.
   \label{chap2-eq-similarity}
\end{equation}
The Hausdorff dimension described previously can be seen as a generalization
of this similarity dimension. Unfortunately, similarity dimension is
only meaningful for a small class of strictly self-similar sets.

\begin{figure}[p]
  \centerline{\epsffile{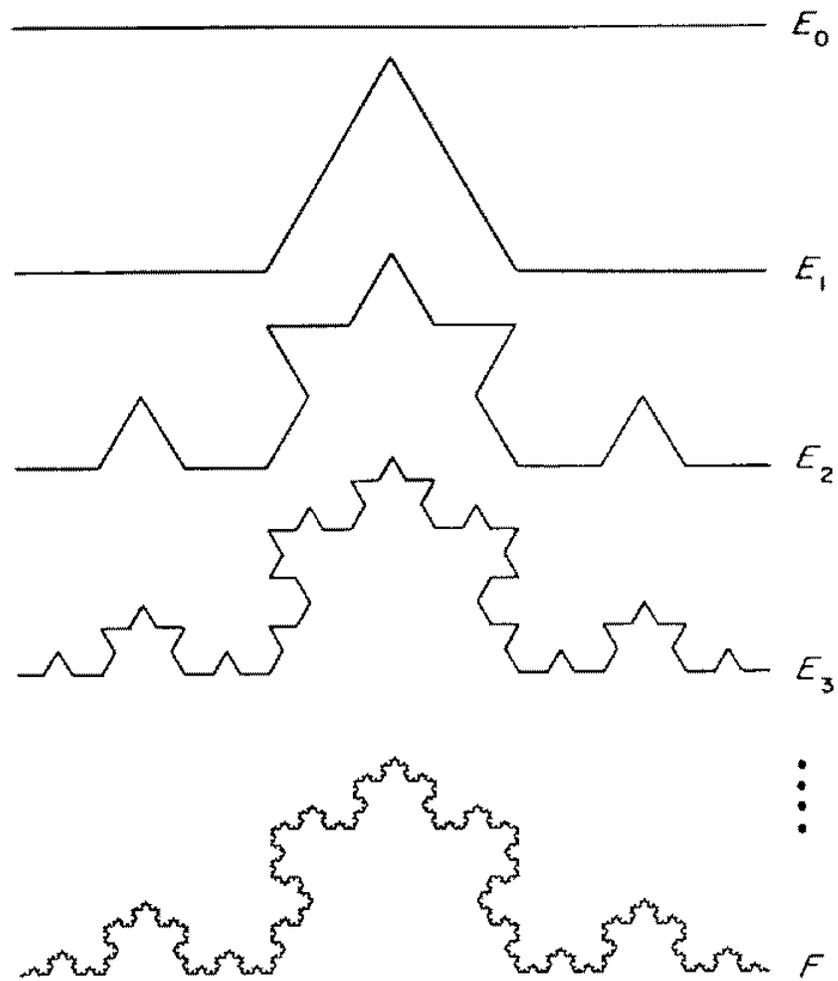}}
  \caption[\sf Construction of the von Koch curve (Falconer 1990).]{\sf
	   Construction of the von Koch curve $F$. At each stage,
	   the middle third of each interval is replaced by the
	   other two sides of an equilateral triangle (Falconer 1990).}
	   \label{2-5}
\end{figure}

Let us see some examples of calculations of the similarity dimension of
sets. Figure \ref{2-5} shows the construction of the von Koch curve, and
any of its segments of unit length is composed of 4 sub-segments each of
which is scaled down by a factor 1/3 from its parent. Therefore, its
similarity dimension is $D=~\log 4~/~\log 3~\cong~1.26$. This
non-integer dimension, greater than one but less than two, reflects the
properties of the curve. It somehow fills more space than a simple line
($D=1$), but less than a Euclidean area of the plane ($D=2$). The figure
\ref{2-5} also shows that the von Koch curve has a finite structure
which is reflected in irregularities at all scales; nonetheless, this
intricate structure stems from a basically simple construction. Whilst
it is reasonable to call it a curve, it is too irregular to have
tangents in the classical sense. A simple calculation on the von Koch
curve shows that $E_k$ is of length $ { \left( \frac{4}{3} \right) }^k$;
letting $k$ tend to infinity implies that $F$ has infinite length. On
the other hand, $F$ occupies zero area in the plane, so neither length
nor area provides a very useful description of the size of $F$.

After this discussion we start to have a better idea of what those
fractal dimensions mean. Roughly, a dimension provides a description of
how much space a set fills. It is a measure of the prominence of the
irregularities of a set when viewed at very small scales. We may regard
the dimension as an index of complexity. We can therefore expect that
a shape with a high dimension will be more complicated than another
shape with a lower dimension.

\begin{figure}[thb]
  \centerline{\epsffile{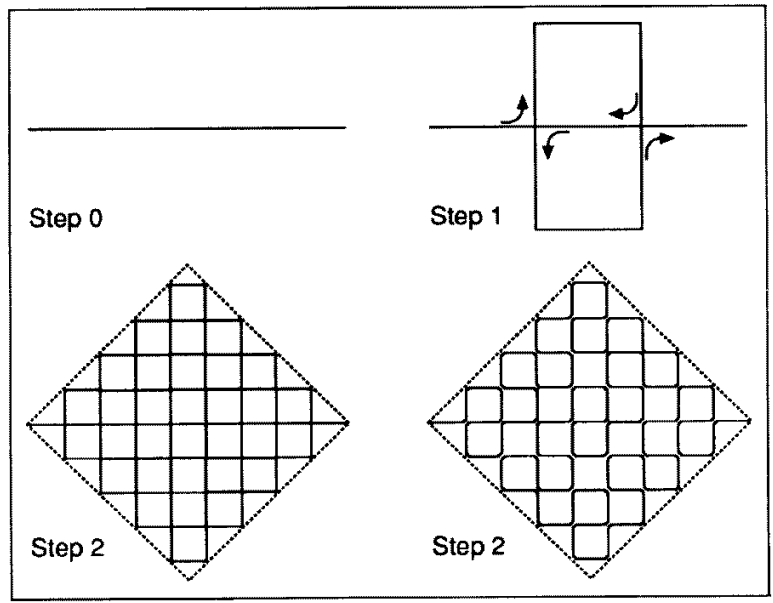}}
  \caption[\sf Construction of Peano's plane-filling curve with
           initiator and generator (Peitgen, J\"{u}rgens and Saupe
	   1992).]{\sf Construction of Peano's plane-filling curve with
           initiator and generator. In each step the  segment is divided
	   in $N=9$ parts, each scaled down by $\delta = 1/3$. This
	   means the similarity dimension is $D=2$. For reasons of
	   clarity the corners where the curve self-contacts have been
	   slightly rounded (Peitgen, J\"{u}rgens and Saupe 1992).}
	   \label{2-6}
\end{figure}

A nice and mathematically fundamental example of this property are the
plane-filling curves mentioned above, which we are now in position to
discuss in a more quantitative manner. There are many different ways of
constructing these
plane-filling curves and here we shall take the original generator 
discussed by Peano. Figure \ref{2-6} shows the construction of the curve
and we can see that the generator self-contacts. The figure only shows
the first two stages, but it is obvious that after a few iterations the
pattern becomes so complex that paper, pencil and our hands turn out to
be rather clumsy and inadequate tools to draw the figure, being unable
to produce anything with much finer detail than that. Therefore,
some sort of computer graphics is then necessary to obtain a detailed
visualization of such curves. The important aspect of this construction
is that the segment is divided in $N=9$ parts, each scaled down by
$\delta = 1/3$, which means the similarity dimension of the Peano curve
is $D=2$.

So we see that this curve has similarity dimension equals to
the square, although it is a line which does not self-intersect. The
removal of one single point cuts the curve in two pieces which means
that its topological dimension is one. The self-contact points of the
Peano curve are inevitable from a logical and intuitive point of view,
and after an infinite number of iterations we effectively have a way of
mapping a part of a plane by means of a topologically one-dimensional
curve: given some patch of the plane, there is a curve which meets every
point in that patch. The set becomes everywhere dense. The Peano curve
is perfectly self-similar, which is shown very clearly in the figure.
The generator used to construct the Peano curve is not the only 
possible one, and in fact there are many different ways of constructing such
plane-filling curves using different generators, although they are
generically known as Peano curves. Mandelbrot (1983) shows many
different and beautiful examples of different generators of Peano
curves.

\subsubsection{Coastlines}\label{Coastlines}

The Hausdorff and similarity dimensions defined so far provide
definitions of fractal dimension for pure fractals, that is, classical
fractal sets in a mathematically idealized way. Although some of these
classical fractals can be used to model physical structures, what is
necessary now is to discuss real fractal shapes which are encountered in
natural phenomena. Hence, we need to apply as far as possible the
mathematical concepts and tools developed so far in the study of real
fractal structures, and when the mathematical tools are found to be
inadequate or insufficient, we need to develop new ones to tackle the 
problem under study. If the existing tools prove inadequate, it is
very likely that a specific definition of fractal dimension appropriate
to the problem under consideration will have to be introduced. Let us
see next two examples where we actually encounter such situations. 
The first is the classical problem of the length of coastlines, where
a practical application of the equation (\ref{eq-ch2-1}) gives the
necessary tools to characterize the shape.

The length of coastlines is the classical fractal problem that was
characterized as such and solved by Mandelbrot. It provides a very
interesting example of a fractal shape in nature and how our intuitive
notion of rectifiable curves is rather slippery. What we shall attempt
to show next is a brief and straightforward presentation of the
problem and its solution, as nothing replaces the fascinating discussion
made by Mandelbrot himself (Mandelbrot 1983).

\begin{figure}[p]
  \centerline{\epsffile{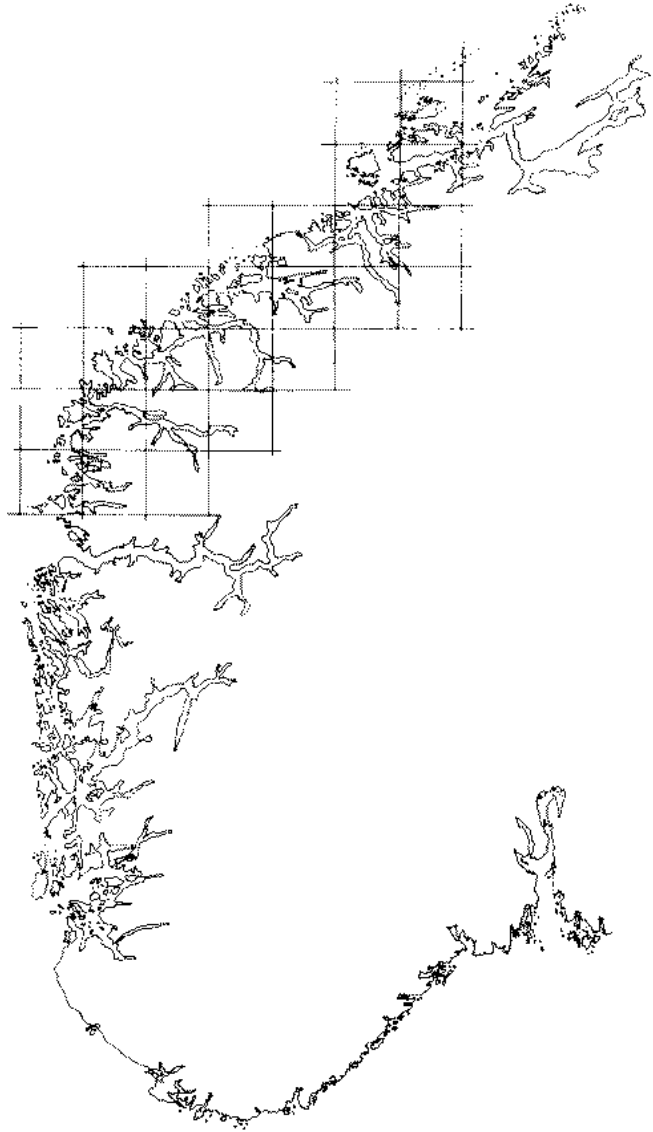}}
  \caption[\sf The coast of the southern part of Norway (Feder 1988).]
	   {\sf The coast of the southern part of Norway. The figure was
           traced from an atlas and digitized at about
	   $1800~\times~1200$ pixels. The square grid indicated has
	   spacing of $\delta \sim 50$ km (Feder 1988).}
	   \label{2-7}
\end{figure}

Let us start with the figure \ref{2-7} showing the southern part of
Norway. The question in this case is: how long is the coast of Norway?
On the scale of the map the deep fjords on the western coast show up
clearly. The details of the southern part are more difficult to resolve,
and if one sails there one finds rocks, islands, bays, faults and gorges
that look much the same but do not show up even in detailed maps. This
fact is absolutely intuitive and anyone who has walked along a beach
and looked at the same beach in the map afterwards can testify it. So
before answering the question of how long is the coast under analysis we
have to decide on whether the coast of the islands should be included.
And what about the rivers? Where does the fjord stop being a fjord and
becomes a river? Sometimes this has an easy answer and sometimes not.
And what about the tides? Are we discussing the length of the coast at
low or high tide?

Despite these initial problems we may press ahead and try to measure the
length of the coast by using an yardstick of length $\delta$ along the
coastline of the map, and count the number of steps $N( \delta )$ needed
to cover it entirely. If we choose a large yardstick then we would not
have to bother about even the deepest fjords, and can estimate the length to be
$L~=~\delta~N(\delta)$. However, somebody could raise objections to
this measurement based on the unanswered questions above, and we could
try a smaller yardstick. This time the large fjords of figure \ref{2-7}
would contribute to the measured length, but the southeastern coast
would still be taken relatively easily in few measurements of the
yardstick. Nevertheless, a serious discussion would demand more detailed
maps, which in consequence reveal more details of the coast, meaning
that a smaller yardstick is then made necessary. Clearly there is no end
to this line of investigation and the problem becomes somewhat
ridiculous. The coastline is as long as we want to make it. It is a
nonrectifiable curve and, therefore, length is an inadequate concept to
compare different coastlines as this measurement is not objective, that
is, it depends on the yardstick chosen.

\begin{figure}[thb]
  \centerline{\epsffile{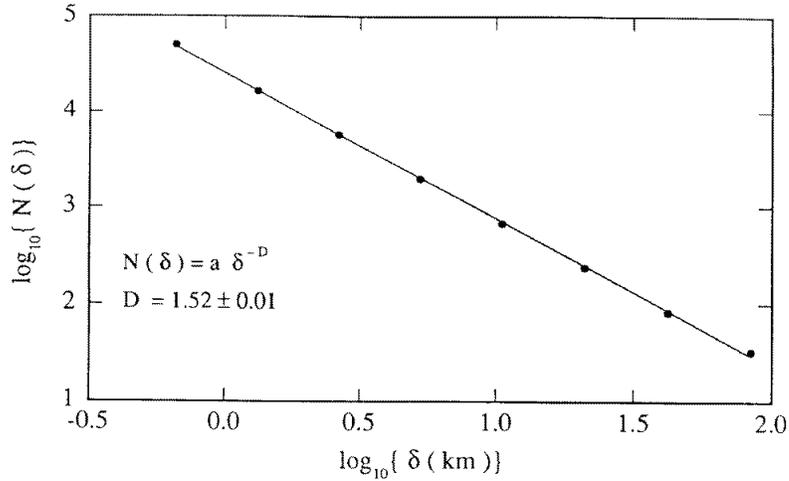}}
  \caption[\sf Covering the coast of Norway with boxes. The fractal
           dimension found is $D~\approx~1.52$ (Feder 1988).]{\sf The
	   number of `boxes' of size $\delta$ needed to cover
           the coastline in the previous figure as a function of $\delta$.
	   The straight line is a fit of
	   $N(\delta)~\propto~\delta^{-D}$ to the observations. The
	   fractal dimension is $D~\approx~1.52$ (Feder 1988).}
	   \label{2-8}
\end{figure}

Coasts, which are obviously self-similar on nature, can, nevertheless,
be characterized if we use a different method based on a practical use
of equation (\ref{eq-ch2-1}). In figure \ref{2-7} the coast of Norway
has been covered with a set of squares with edge length $\delta$, with
the unit of length taken to equal the edge of the frame. Counting the
number of squares needed to cover the coastline gives the number 
$N(\delta)$. If we proceed and find $N(\delta)$ for smaller values
of $\delta$ we are able to plot a graph of $N(\delta)$ versus
$\delta$, for different grid sizes. Now it follows from equation
(\ref{eq-ch2-1}) that asymptotically in the limit of small $\delta$ the
following equation is valid:
\begin{equation}
 N( \delta ) \propto \frac{1}{\delta^D}.
 \label{eq-ch2-3}
\end{equation}
So the fractal dimension $D$ of the coastline can be determined by
finding the slope of $\log~N(\delta)$ plotted as a function of
$\log~\delta$. The resulting plot for the coastline shown in figure
\ref{2-7} is presented in figure \ref{2-8}, with $D~\approx~1.52$, and
the same method produces $D~\approx~1.31$ for the coast of Britain.
These two values agree with the intuition already associated with
fractal dimensions in the sense that the more irregular coast, the
Norwegian in this case, should have a higher value for $D$ than the
British coast. By means of equation (\ref{eq-ch2-3}) we can obtain the
expression for the length of coastlines as
\[
 L \propto \delta^{1-D},
\]
which shows its explicit dependence on the yardstick chosen.
The dimension $D$ in equation (\ref{eq-ch2-3}) is determined by counting
the number of boxes needed to cover the set as a function of the box
size. It is called {\it box counting dimension} or {\it box dimension}.

The box counting dimension proposes a systematic measurement which
applies to any structure in the plane, and can be readily adapted for
structures in the space. It is perhaps the most commonly used method of
calculating dimensions and its dominance lies in the easy and automatic
computability provided by the method, as it is straightforward to count
boxes and maintain statistics allowing dimension calculation. The
program can be carried out for shapes with and without self-similarity
and, moreover, the objects may be embedded in higher dimensional spaces.

\subsubsection{Cluster Dimension}\label{Cluster Dim}

Let us see now the application of fractal ideas to the problem of
aggregation of fine particles, such as those of soot, where an
appropriate fractal dimension has to be introduced. This kind
of application of fractal concepts to real physical systems is under
vigorous development at present, inasmuch as it can be used to study such
systems even in a laboratory, where they can be grown, or in computer
simulations.

The Hausdorff dimension $D$ in equation (\ref{eq-ch2-1}) requires the
size $\delta$ of the covering sets to vanish, but as physical systems in
general have a characteristic lower length scale, we need to take that
into consideration in our physical applications of fractals. For
instance, the problem of the length of coastlines {\it necessarily} involves
a lower cutoff in its analysis as below certain scale, say, at the
molecular level, we are no longer talking about coastlines. For the same
reason we sometimes have to assume upper cutoffs to the fractal
structures we are analysing. This highlights once more an important aspect of
application of fractals to real physical phenomena: each problem must be
carefully analysed not only to look for the appropriate fractal dimension
(or dimensions as we may have more than one defined in the problem), but
also to see to what extent the fractal hypothesis is valid to the case
in study.

\begin{figure}[thb]
  \centerline{\epsffile{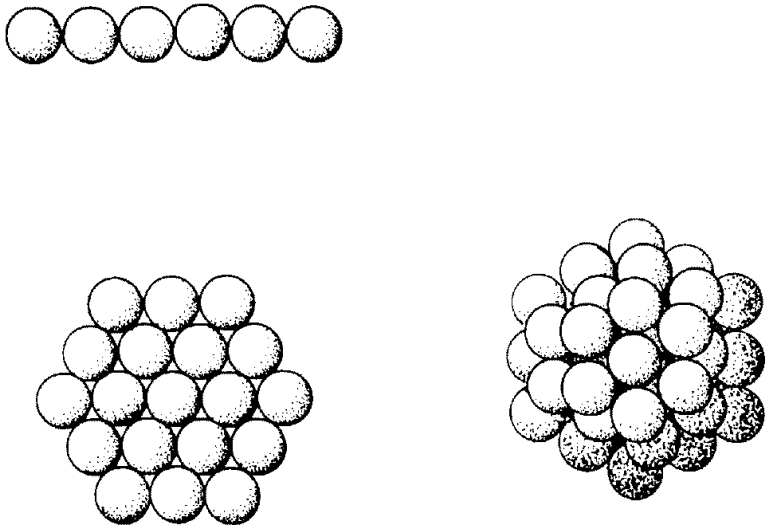}}
  \caption{\sf Simple aggregation of spherical monomers (Feder 1988).}
	   \label{2-9}
\end{figure}

Let us see a specific example. In order to apply the ideas of the
previous sections, we can replace a mathematical line by a linear chain
of ``molecules'' or monomers. Figure \ref{2-9} shows the replacement of
a line by a chain of monomers, a two-dimensional set of points by a
planar collection of monomers, and a volume by a packing of spheres. Let
us call the radius $R_0$ the smallest length scale of the structure
under study. In this case $R_0$ will be the radius of the monomers in
figure \ref{2-9}. The number of monomers in a chain of length $L=2R$ is
\[
  N={\left( \frac{R}{R_0} \right) }^1.
\]
For a group of monomers in a circular disk we have the proportionality
\[
  N \propto {\left( \frac{R}{R_0} \right) }^2.
\]
For the three-dimensional close packing of spherical monomers into a
spherical region of radius $R$, the number of monomers is
\[
  N \propto {\left( \frac{R}{R_0} \right) }^3.
\]
By generalising these relations we may say that the number of particles
and the cluster size measured by the smallest sphere of radius $R$
containing the cluster is given by
\begin{equation}
 N \propto {\left( \frac{R}{R_0} \right) }^D,
  \label{eq-ch2-4}
\end{equation}
where $D$ is the dimension of the distribution that may be non-integer,
that is, fractal. Equation (\ref{eq-ch2-4}) is a {\it number-radius
relation} and $D$ is the {\it cluster fractal dimension} of the aggregation.
The cluster fractal dimension is a measure of how the cluster fills the
space it occupies.

A fractal cluster has the property of having a decreasing average
density as the cluster size increases, in a way described by the
exponent in the number radius relation. The average density will have
the form
\begin{equation}
 \rho(R) \sim {R_0}^{-D} R^{D-E},
 \label{eq-ch2-5}
\end{equation}
where $E$ is the Euclidean dimension of the space where the cluster is
placed. Therefore, a cluster is not necessarily fractal, even if it
is porous or formed at random, as its density may be constant. Note that
the shape of the cluster is {\it not} characterized by the cluster
fractal dimension, although $D$ does characterize, in a quantitative way,
the cluster's feature of ``filling'' the space.

\begin{figure}[thb]
  \centerline{\epsffile{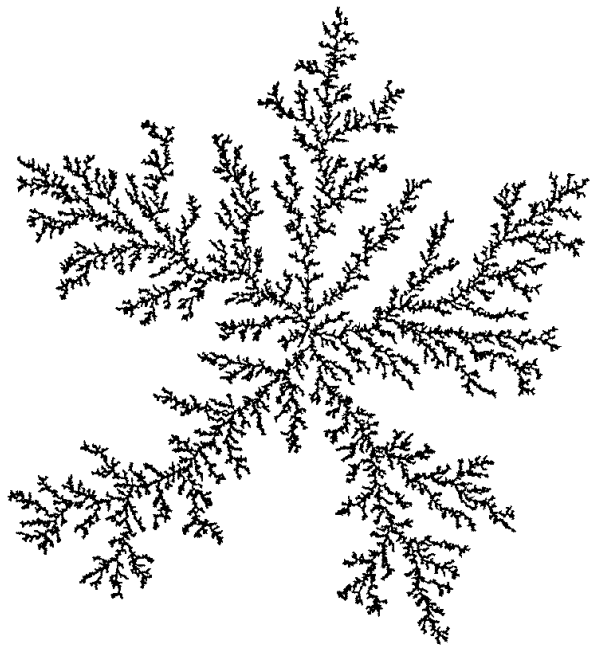}}
  \caption[\sf Random cluster obtained from a two-dimensional DLA
           process (Feder 1988).]{\sf Random cluster containing
	   50,000 particles obtained from a two-dimensional
           diffusion-limited aggregation process (DLA)
	   with $D~=~1.71$ (Feder 1988).}
	   \label{2-10}
\end{figure}

Figure \ref{2-10} shows a very much studied type of cluster, obtained
by the {\it diffusion-limited aggregation process} (DLA). In this
process the cluster is start\-ed by a seed in the centre, and wandering
monomers stick to the growing cluster when they reach it: if the random
walker contacts the cluster, then it is added to it and another walker
is released at a random position on the circle. This type of aggregation
process produces clusters that have a fractal dimension $D~=~1.71$ for
diffusion in the plane. Numerical simulations show that the fractal
dimension is $D~=~2.50$ for clusters in three-dimensional space.

Before closing this section, it seems appropriate to discuss something
about the limitations of the fractal geometry. For this purpose we shall
quote a paragraph from Peitgen, J\"{u}rgens and Saupe (1992, p. 244)
which beautifully expresses this point.

\begin{quotation}
 \small

 ``The concept of fractal dimension has inspired scientists to a host
 of interesting new work and fascinating speculations. Indeed, for a
 while it seemed as if the fractal dimensions would allow us to
 discover a new order in the world of complex phenomena and structures.
 This hope, however, has been dampened by some severe limitations. For
 one thing, there are several different dimensions which give different
 answers. We can also imagine that a structure is a mixture of different
 fractals, each one with a different value of box counting dimension. In
 such case the conglomerate will have a dimension which is simply the
 dimension of the component(s) with the largest dimension. That means
 the resulting number cannot be characteristic for the mixture. What we
 would really like to have is something more like a spectrum of numbers
 which gives information about the distribution of fractal dimensions in
 a structure. This program has, in fact, been carried out and runs under
 the theme {\it multifractals}.''
\end{quotation}
For reasons of simplicity we shall not deal with multifractals in this paper.

As a final remark, as Pietronero (1988) has pointed out, the fractal
concept provides a {\it description} of these irregular structures on
nature, but it does not imply the formulation of a theory for them.
Indeed, Mandelbrot has not produced a theory to explain how these
structures actually arise from physical laws. A study of the
interrelations between fractal geometry and physical phenomena is what
may be called the ``theory of fractals'', and forms the objective of
intense activity in the field nowadays. This activity is basically
divided in two main streams. The first tries to understand how it has 
come about that many shapes in nature present fractal properties. Hence,
the basic question to answer is: where do fractals come from? The second
approach is to assume as a matter of fact the existence of fractal
structures and to study their physical properties. This generally
consists of assuming a simple fractal model as a starting point and
studying, for example, some basic physical property, like the diffusion
on this structure.

As examples of such studies, there are the {\it fractal growth models},
which are based on a stochastic growth process in which the probability
is defined through Laplace equation ({\it e.g.}, DLA process). They are
considered the prototypes of many physical phenomena that generate fractal
structures. As other examples, we have the {\it self-organized critical
systems}, which are such that a state with critical properties is reached
spontaneously, by means of an irreversible dynamical evolution of a complex
system ({\it e.g.}, sandpile models). They pose problems similar to
the fractal growth process, and use theoretical methods inspired by the
latter, like the so-called ``fixed scale transformation'', that allows to
deal with irreversible dynamics of these process and to calculate
analytically the fractal dimension.

\section{The Fractal Hypothesis for the Distribution of Galaxies}\label{Distr
Galaxies}

In this section we shall discuss how the fractal concept can be used to
study the large scale distribution of galaxies in the observed
universe. We start with a brief summary of the standard methods used to
study this distribution. Later we will see the problems of this
orthodox analysis and the answers given to these problems when we
assume that the large scale distribution of galaxies forms a
self-similar fractal system. Some implications of the use of fractal ideas to 
describe the distribution of galaxies, like a possible crossover to
homogeneity, are also presented.  

\subsection{The Standard Correlation Function Analysis}\label{Standard
Analysis}

The standard statistical analysis assumes that the objects under
discussion (galaxies) can be regarded as point particles that are
distributed homogeneously on a sufficiently large scale. This means that
we can meaningfully assign an average number density to the
distribution and, therefore, we can characterize the galaxy distribution
in terms of the extent of the departures from uniformity on various
scales. The correlation function as introduced in this field by P. J. E.
Peebles around 25 years ago is basically the statistical tool that
permits the quantitative study of this departure from homogeneity.

We shall discuss in a moment how to obtain the explicit form of the
two-point correlation function, but it is important to  point out right now
two essential aspects of this method. First that this analysis fits very
well in the standard Friedmannian cosmology which {\it assumes} spatial
homogeneity, but it does not take into consideration any effect due to
the curvature of the spacetime. In fact, this method neglects this
problem altogether under the assumption that the scales under study are
relatively small, although it does not offer an answer to the question
of where we need to start worrying about the curvature effects. In
other words, this analysis does not tell us what scales can no longer
be considered relatively small.

Secondly that if the homogeneity assumption is not justified this
analysis is {\it inapplicable}. Moreover, because this analysis starts
by assuming the homogeneity of the distribution, {\it it does not offer
any kind of test for the hypothesis itself}. In other words, {\it this
correlation analysis cannot disprove the homogeneous hypothesis}.

Let us present now a straightforward discussion on how this statistic
can be derived (Raine 1981). If the average number
density of galaxies is $\bar{n} = N/V$ then we have to go on average a
distance ${ ( \bar{n} )}^{-1/3}$ from a given galaxy before another is
encountered. This means that local departures from uniformity can be
described if we specify the distance we actually go from any particular
galaxy before encountering another. This will sometimes be larger than
average, but sometimes less. Specifying this distance in each case is
equivalent of giving the locations of all galaxies. This is an awkward
way of doing things and does not solve the problem. What we require is
a statistical description giving the probability of finding the nearest
neighbour galaxy within a certain distance.

However, as the probability of finding a galaxy closer than, say, 50 kpc
to the Milky Way is zero, and within a distance greater than this value
is one, this is clearly not the sort of probability information we are
after; what is necessary is some sort of average. Now we can think that the
actual universe is a particular realization of some statistical
distribution of galaxies, some statistical law, and the departure from
randomness due to clustering will be represented by the fact that the
average value over the statistical ensemble of this separation is less than
${ ( \bar{n} ) }^{-1/3}$.

In practice, however, we do not have a statistical ensemble from which
the average value can be derived, so what we can do is to take a spatial
average over the visible universe, or as much of it as has been
catalogued, in place of an ensemble average. This only makes sense if the
departure from homogeneity occurs on a scale smaller than the depth of
the sample, so that the sample will statistically reflect  the
properties of the universe as a whole. In other words, we need to
achieve a {\it fair sample} of the whole universe in order to fulfill this
program, and this fair sample ought to be homogeneous, by
assumption. If, for some reason, this fair sample is not achieved, or is not
achievable, that is, if the universe has no meaningful average density,
this whole program breaks down.

For a completely random but homogeneous distribution of galaxies, the
probability $dP_1$ of finding a galaxy in an infinitesimal volume
$dV_1$ is proportional to $dV_1$ and to $\bar{n}$, and is independent of
position. So we have
\[
  dP_1 = \frac{ \bar{n} }{N} dV_1,
\]
where $N$ is the total number of galaxies in the sample. The meaning of
the probability here is an average over the galaxy sample; the space is
divided into volumes $dV_1$ and we count the ratio of those cells which
contain a galaxy to the total number. The probability of finding two
galaxies in a cell is of order ${(dV_1)}^2$, and so can be ignored in
the limit $dV_1 \to 0$. It is important to state once more that this
procedure only makes sense if the galaxies are distributed randomly
on some scale less than that of the sample.

Suppose now that the galaxies were not clustered. In that case the
probability $dP_{12}$ of finding galaxies in volumes $dV_1$ and $dV_2$
is just the product $dP_1 dP_2$ of probabilities of finding each of the
galaxies, since in a random distribution the positions of galaxies are
uncorrelated. Now, if the galaxies were correlated we would have a
departure from the random distribution and, therefore, the joint
probability will differ from a simple product. The {\it two-point
correlation function} $\xi (\vec{r}_1, \vec{r}_2)$ is by definition a
function which determines the difference from a random distribution. So
we put
\begin{equation}
 dP_{12} = { \left( \frac{ \bar{n} }{N} \right) }^2
	   \left[ 1+ \xi (\vec{r}_1, \vec{r}_2) \right] dV_1 dV_2
 \label{eq-ch2-6}
\end{equation}
as the expression of finding a pair of galaxies in volumes $dV_1$,
$dV_2$ at positions $\vec{r}_1$, $\vec{r}_2$. Obviously, the assumption
of randomness on sufficiently large scales means that $\xi (\vec{r}_1,
\vec{r}_2)$ must tend to zero if $\left| \vec{r}_1 - \vec{r}_2 \right|$
is sufficiently large. In addition, the assumption of homogeneity means
that $\xi$ cannot depend on the location of the galaxy pair, but only
on the distance $\left| \vec{r}_1 - \vec{r}_2 \right|$ that separates
them, as the probability must be independent of the location of the
first galaxy. If $\xi$ is positive we have an excess probability over
a random distribution and, therefore, clustering. If $\xi$ is negative
we have anti-clustering. Obviously $\xi > -1$. The two-point
correlation function can be extended to define $n$-point correlation
functions, which are functions of $n-1$ relative distances, but in
practice computations have not been carried out beyond the four-point
correlation function.

It is common practice to replace the description above using point
particles by a continuum description. So if galaxies are thought to be
the constituent parts of a fluid with variable density $n(\vec{r})$, and
if the averaging over a volume $V$ is carried out over scales large
compared to the scale of clustering, we have
\begin{equation}
 \frac{1}{V} \int_V n(\vec{r}) dV = \bar{n},
 \label{eq-ch2-6A}
\end{equation}
where $dV$ is an element of volume at $\vec{r}$. The joint probability
of finding a galaxy in $dV_1$ at $\vec{r} + \vec{r}_1$ and in $dV_2$ at
$\vec{r} + \vec{r}_2$ is given by
\[
  { \left( \frac{1}{N} \right) }^2 n(\vec{r} + \vec{r}_1)
  n(\vec{r} + \vec{r}_2) dV_1 dV_2.
\]
Averaging this equation over the sample gives
\[
  dP_{12} =  \frac{1}{N^2 V} \int_V  n(\vec{r} + \vec{r}_1) n(\vec{r} +
  \vec{r}_2) dV dV_1 dV_2.
\]
Now if we compare the equation above with equation (\ref{eq-ch2-6}) we
obtain
\begin{equation}
   { \bar{n}}^2 \left[ 1+ \xi(\vec{\tau}) \right] = \frac{1}{V} \int_V
   n(\vec{R}) n(\vec{R}+\vec{\tau}) dV,
 \label{eq-ch2-7}
\end{equation}
where $\vec{\tau} = \vec{r}_2 - \vec{r}_1$, $\vec{R} = \vec{r}
+ \vec{r}_1$ and $dV$ is the volume element at $\vec{R}$. 

Related to the correlation function is the so-called {\it power spectrum}
of the distribution, defined by the Fourier transform of the correlation
function. It is also possible to define an {\it angular correlation
function} which will express the probability of finding a pair of galaxies
separated by a certain angle, and this is the function appropriate to
studying catalogues of galaxies which contain only information on the
positions of galaxies on the celestial sphere, that is, to studying the
projected galactic distribution when the galaxy distances are not
available. Further details about these two functions can be found, for
instance, in Raine (1981, p. 10).
Finally, for the sake of easy comparison with other works it is
useful to write equation (\ref{eq-ch2-7}) in a slightly different notation:
\begin{equation}
 \xi(r) = \frac{ \langle n(\vec{r}_0) n(\vec{r}_0+\vec{r}) \rangle }
	       { { \langle n \rangle }^2} - 1.
 \label{eq-ch2-9}
\end{equation}

The usual interpretation of the correlation function obtained from the
data is as follows. When $\xi~\gg~1$ the system is strongly correlated
and for the region when $\xi~\ll~1$ the system has small correlation.
From direct calculations from catalogues it was found that at small
values of $r$ the function $\xi(r)$ can be characterized by a power
law (Pietronero 1987; Davis et al. 1988; Geller 1989):
\begin{equation}
 \xi(r) \approx A r^{- \gamma}, \ \ \ \ (\gamma \approx 1.7),
 \label{eq-ch2-10}
\end{equation}
where $A$ is a constant. This power law behaviour holds for galaxies and
clusters of galaxies. The distance $r_0$ at which $\xi~=~1$ is called
the {\it correlation length}, and this implies that the system becomes
essentially homogeneous for lengths appreciably larger than this
characteristic length. This also implies that there should be no
appreciable overdensities (superclusters) or underdensities (voids)
extending over distances appreciably larger than $r_0$.

\subsection{Difficulties of the Standard Analysis}\label{Difficulties Standard
Analysis}

The first puzzling aspect found using the method just described
is the difference
in the amplitude $A$ of the observed correlation function (\ref{eq-ch2-10})
when measured for galaxies and clusters of galaxies. While the exponent
$\gamma$ is approximately 1.7 in both cases, for galaxies
$A_G~\simeq~20$ and for clusters $A_C~\simeq~360$. Less accurately,
superclusters of galaxies were found to have $A_{SC}~\simeq~1000-1500$
(see Pietronero 1987 and references therein). The correlation length
was found to be $r_0~\simeq~5$~h$^{-1}$ Mpc for galaxies and
$r_0~\simeq~25$~h$^{-1}$ Mpc for clusters.

These are puzzling results, because as $A_C~\simeq~18A_G$, clusters
appear to be much more correlated than galaxies, although they are
themselves made of galaxies. Similarly superclusters will then appear to
be more correlated than clusters. From the interpretation of $\xi(r)$
described above, the galaxy distribution becomes homogeneous at the
distance $\simeq$ 10-15 h$^{-1}$ Mpc where $\xi(r)$ is found to become zero,
while clusters and superclusters are actually observed at much larger
distances, in fact up to the present observational limits.

\begin{figure}[p]
  \centerline{\epsffile{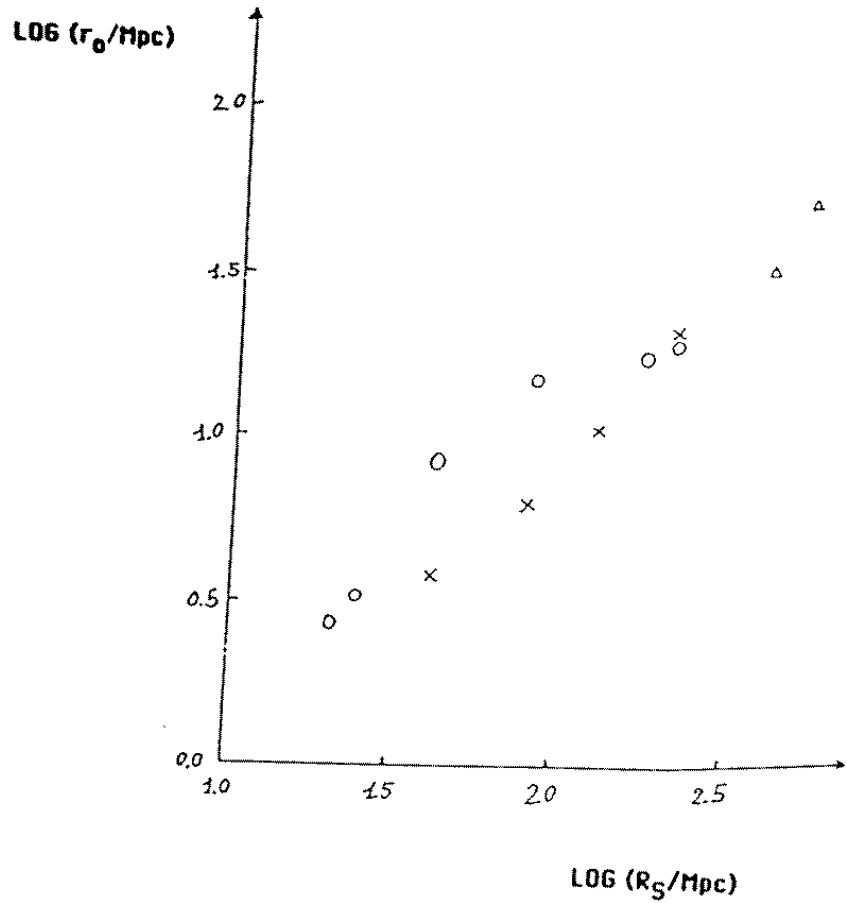}}
  \caption[\sf The increase of the correlation length with the sample
          size found by Einasto, Klypin and Saar (1986).]{\sf The
	  behaviour of $r_0$ plotted against the sample size
          $R_S$ found by Einasto, Klypin and Saar (1986). The different
	  symbols (open circles, crosses and triangles) mean the
	  different directions from where the samples were taken in the sky.
	  (Calzetti et al. 1987). Further extensions of this behaviour, to
	  around 100 Mpc, are shown in Pietronero (1997), fig.\ 2.}
	   \label{2-11}
\end{figure}

The second difficulty of the standard analysis was first found by
Einasto, Klypin and Saar (1986) and later confirmed by Davis et~al. 
(1988) (see also Calzetti et~al. 1987 and Coleman, Pietronero and Sanders
1988, Pietronero 1997). They found that the correlation
length $r_0$ {\it increases with the sample size}. Figure \ref{2-11}
shows this dependence clearly, and it is evident that this result brings
into question the notion of a universal galaxy correlation function.

\begin{figure}[p]
  \centerline{\epsffile{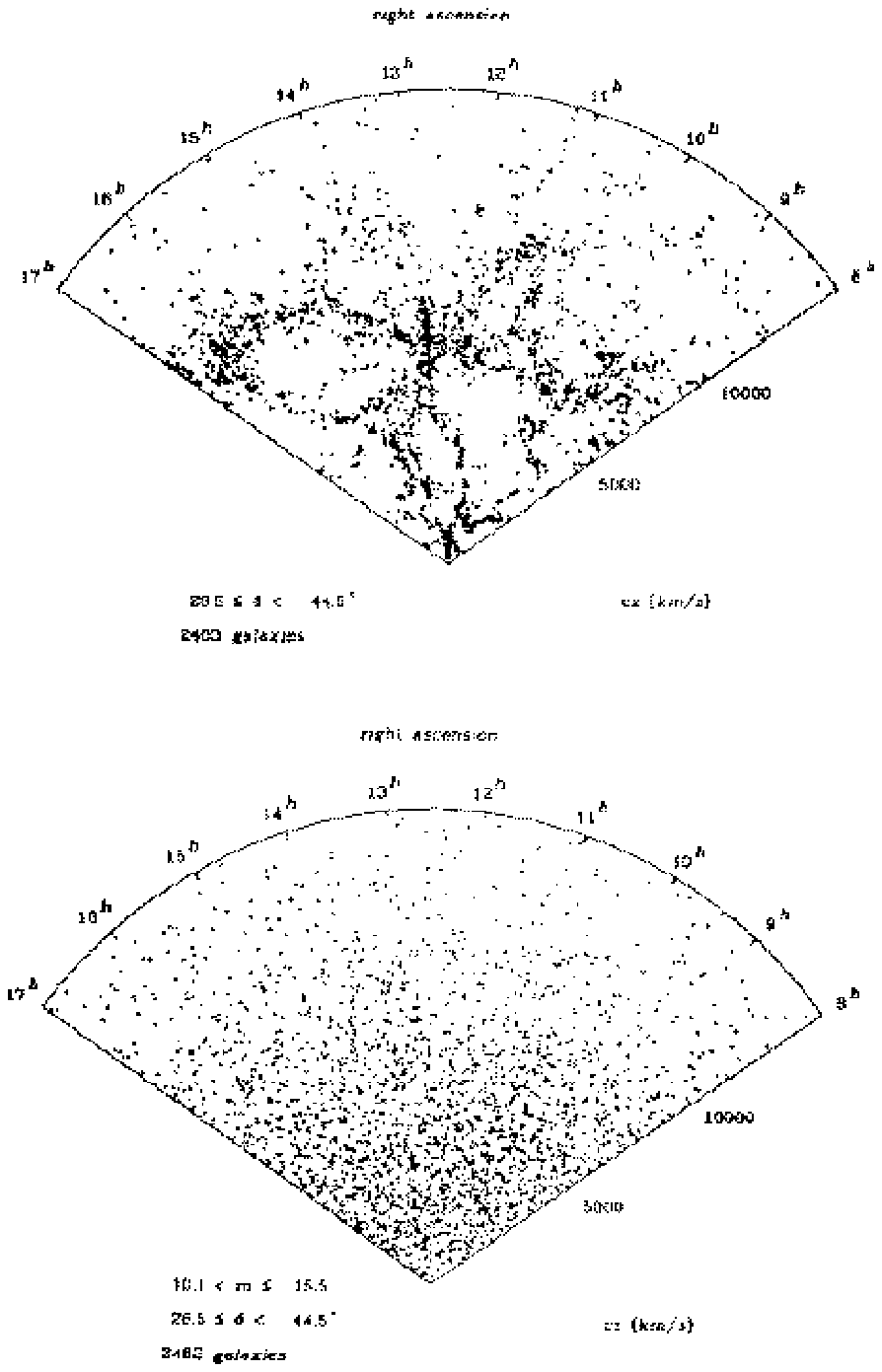}}
   \caption[\sf The observed distribution of galaxies in the
	   $18^{~\mbox{o}}$ wide conic slice compared with a sample of
	   randomly distributed points (Geller 1989).]{\sf The picture
	   on top shows the observed distribution of
           galaxies in the $18^{ \mbox{o}}$ wide slice centered at
	   $35.5^{ \mbox{o}}$. Voids and clusters are clearly visible as
	   well as the lack of homogenization of the sample. The largest
	   inhomogeneities are comparable with the size of the sample
	   and, therefore, it is not large enough to be considered {\it fair}. 
           The picture below shows a sample of 2483 randomly distributed 
           points (Geller 1989).}
          \label{2-12}
\end{figure}

The third problem of the standard analysis has to do with the
homogeneity assumption itself and the possibility of achieving a 
fair sample, which should not be confused with a homogeneous sample as 
the standard analysis usually does. A fair sample is one in which there exists 
enough points from where we are able to derive some unambiguous statistical 
properties. Therefore {\it homogeneity must be regarded as a property of the 
sample and not a condition of its statistical validity}. Improvements in 
astronomical detection techniques, in particular the new sensors and 
automation, enabled astronomers to obtain a large amount of galaxy redshift 
measurements per night, and made it possible by the mid 1980's to map the 
distribution of galaxies in three dimensions. The picture that emerged from 
these surveys was far from the expected homogeneity: clusters of galaxies,
voids and superclusters appeared in all scales, with no clear homogenization
of the distribution. The first ``slice'' of the universe shown by de Lapparent, 
Geller and Huchra (1986) confirmed this inhomogeneity with very clear pictures. 
More striking is the comparison of these observed slices with a randomly
generated distribution where the lack of homogenization of the observed
samples is clear (see figure \ref{2-12}). Deeper surveys (Saunders et~al.\
1991) show no sign of any homogeneity being achieved so far, with the same
self-similar structures still being identified in the samples.


Those problems together with the power law behaviour of $\xi(r)$ clearly
call for an explanation, and while many have been proposed they usually
deal with each of these issues separately. As we shall see, the
fractal hypothesis, on the other hand, deals with all these problems
as a whole and offers an explanation to each of them within the fractal
picture. While we do not intend to claim that the fractal hypothesis is
the only possible explanation to these problems, whether considering them
together or separately, from now on in this paper we shall take
the point of view that fractals offer an attractively simple
description of the large scale
distribution of galaxies and that the model offered by them deserves a
deep, serious and unprejudiced investigation, either in a Newtonian
or relativistic framework. From its basis, the fractal hypothesis in many
ways represents a radical departure from the orthodox traditional view
of an {\it observationally} homogeneous universe, which is challenged from 
its very foundations in many respects.

\subsection{Correlation Analysis without Assumptions}\label{Corr Anal without
Assump}

Before we discuss the fractal model itself, it is convenient to look
first at the statistical tools where a correlation appropriate to a
fractal distribution can be derived. The method that is going to be
introduced obviously does not imply the existence of a fractal
distribution, but if it exists, it is able to describe it properly (see Pietronero 1987; 
Coleman and Pietronero 1992; Pietronero, Montuori and Sylos Labini 1997; Sylos  Labini, Montuori and Pietronero 1998). The 
appropriate analysis of pair correlations should therefore be performed using 
methods that can check homogeneity or fractal properties without assuming 
{\it a priori} either one. This is not the case of the function
$\xi (r)$, which is based on {\it a priori} and untested assumption of
homogeneity. For this purpose it is useful to start with a basic discussion
on the concept of correlation. 

\begin{figure}[htb]
  \centerline{\epsffile{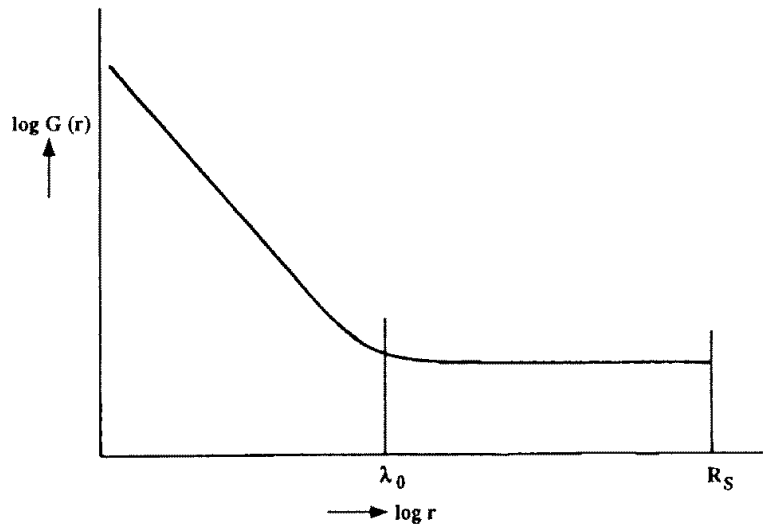}}
  \caption[\sf Schematic representation of the galaxy correlation with a
           crossover to homogeneity (Coleman and Pietronero~1992).]{\sf This
	   illustration represents schematically the 
	   expected behaviour for the
           galaxy correlation density of a correlated system with a
	   crossover to homogeneity. The flat behaviour of the function
	   $G(r)$ beyond some correlation length $\lambda_0$ corresponds
	   to the unambiguous sign of homogeneity \ (Coleman and
	   Pietronero~1992).}
	  \label{2-14}
\end{figure}

If the presence of an object at $r_1$ influences the probability of finding
another object at $r_2$ these points are said to be correlated. So there
is correlation at a 
distance $\vec{r}$ from $\vec{r}_0$ if, on average,
\[
 G(r) = \langle n(\vec{r}_0) n(\vec{r}_0+\vec{r}) \rangle \not= 
	{ \langle n \rangle }^2.
\]
On the other hand there is no correlation if
\[
 G(r) \approx { \langle n \rangle }^2.
\]
From this it is clear that non-trivial structures like voids or
superclusters are made, by definition, from correlated points. Hence,
the correlation length definition will only be meaningful if it
separates correlated regions from uncorrelated ones. Figure \ref{2-14}
shows what would be the typical behaviour we would expect from the
galaxy correlation if there is an upper cutoff to homogeneity. The
power law decay will be eventually followed by a flat region that
corresponds to the homogeneous distribution. In this case the
correlation length $\lambda_0$ is the distance at which there is a
change in the correlation $G(r)$: it changes from a power law behaviour
to a homogeneous regime and the average density becomes the same, being
independent of the position.

Actually in figure \ref{2-14} we have a situation where the sample size
$R_S$ is larger than $\lambda_0$. If we had $R_S < \lambda_0$, then the
average density measured would correspond to an overdensity, different
from the real average density of the distribution. The precise value of
this overdensity would then depend explicitly on the sample radius, and
in this case the function $\xi (r)$ becomes spurious because the
normalization of correlated density by the average density
(eq.~\ref{eq-ch2-9}) will depend explicitly on $R_S$. Only in the limiting
case where $R_S~\gg~\lambda_0$ will the length $r_0$ indeed be related
to the correlation length $\lambda_0$~\footnote{ \ See \S \ref{Cross Homo} 
for the proof of this statement.}. This is in fact the case for
liquids, where the two-point correlation function $\xi (r)$ was
originally introduced, as any reasonable sample size larger than some
atomic scale for, say, water, will contain so many molecules that
its average density is a well defined physical property. Because water
has a well defined value for $\langle n \rangle$, the function $\xi(r)$
will be the same for a glass of water and for a lake, referring only to
the physical properties of water and not to the size of the glass.
However, by just looking at figure \ref{2-12} it is quite clear
that this is not the case for the distribution of galaxies. As pointed
out by Geller (1989), at least up to the present observational limits
the galaxy average density is not well defined.

The function $G(r)$ as defined above has, however, the factor
$\langle n \rangle~=~N/V$ which may, in principle, be an explicit
function of the sample's size scale. Therefore, it is appropriate to
define the {\it conditional density} $\Gamma(r)$ as,
\begin{equation}
 \Gamma (r) \equiv \frac{ \langle n(\vec{r}_0) n(\vec{r}_0 + \vec{r}) \rangle}
	      { \langle n \rangle}.
 \label{eq-Gamma-1}
\end{equation}
By the definition of the average number density of galaxies, we have
\begin{equation}
 \langle n(\vec{r}_0) n(\vec{r}_0 + \vec{r}) \rangle =
   \frac{1}{V} \int_V n(\vec{r}_0) n(\vec{r}_0 + \vec{r}) dV,
 \label{eq-avenumgal}
\end{equation}
and together with $\langle n \rangle~=~N/V$, equation (\ref{eq-Gamma-1}) 
becomes
\begin{equation}
 \Gamma (r) = \frac{1}{N} \int_V n(\vec{r}_0) n(\vec{r}_0 + \vec{r}) dV.
 \label{eq-Gamma-2}
\end{equation}
Assuming
\[
 n(\vec{r}) = \sum^N_{i=1} \delta (\vec{r} - \vec{r}_i),
\]
and remembering that
\[
 \int_{-\infty}^{+\infty} \delta (x-y) \varphi (y) dy = \varphi (x), \hspace{5mm}
 \delta (x) = \delta (-x),
\]
equation (\ref{eq-Gamma-2}) may be rewritten as
\begin{equation}
 \Gamma (r) = \frac{1}{N} \sum^N_{i=1} n(\vec{r}_i + \vec{r})
	        = {\langle n(\vec{r}_i + \vec{r}) \rangle}_i,
  \label{eq-ch2-2-11a}
\end{equation}
where $n(\vec{r}_i~+~\vec{r})$ is the conditional density of the
{\it i}th object, and ${\langle n(\vec{r}_i~+~\vec{r})\rangle}_i$ the final
average that refers to all occupied points $r_i$. 
This corresponds to assigning an unit mass to all points occupied by
galaxies, with the {\it i}th galaxy at $\vec{r}_i$ and $N$ being the
total number of galaxies. Equation (\ref{eq-ch2-2-11a}) measures the
average density at distance $r$ from an occupied point, and the volume
$V$ in that equation is purely nominal and should be such to include all
objects, but it does not appear explicitly in equation
(\ref{eq-ch2-2-11a}). $\Gamma(r)$ is more convenient than $G(r)$ for
comparing different catalogues since the
size of the catalogue only appears via the combination
$\frac{1}{N}~\sum^N_{i=1}$ and this means that a larger sample volume
will only increases the number of objects $N$. Hence a larger sample
size implies a better statistic.

Equations (\ref{eq-ch2-9}) and (\ref{eq-ch2-2-11a}) are simply related by
\begin{equation}
 \xi(r) = \frac{\Gamma (r) }{ \langle n \rangle } -1.
 \label{eq-ch2-2-11b}
\end{equation}
We can also define the {\it conditional average density}
\begin{equation}
 \Gamma^{\ast} (r) = \frac{1}{V} \int_V \Gamma (r) dV,
 \label{eq-ch2-2-11c}
\end{equation} 
which gives the behaviour of the average density of a sphere of radius
$r$ centered around an occupied point averaged over all occupied points 
\footnote{ \ See Coleman and Pietronero (1992) for a more detailed
discussion on this subject with many examples of calculations of
$\Gamma(r)$, $\Gamma^{\ast}(r)$ and $\xi(r)$ using test samples.}, 
and the {\it integrated conditional density}
\begin{equation}
 I(r) = 4 \pi \int_{0}^{r}{r^\prime }^2\Gamma (r^\prime) d{r^\prime },             
 \label{eq-ch2-2-11d}
\end{equation} 
which is the number of galaxies of a spherical region of radius r.

\begin{figure}[p]
  \centerline{\epsffile{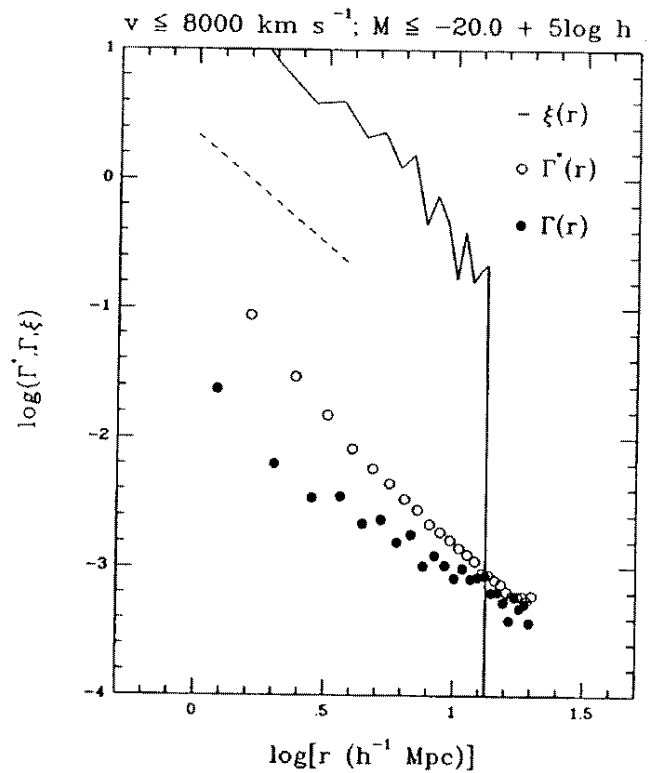}}
  \caption[\sf Statistical tools without assumptions show no
                     homogenization of the CfA redshift survey (Coleman,
	  Pietronero and Sanders~1988).]{\sf $\Gamma(r)$,
	  $\Gamma^{\ast}(r)$ and $\xi(r)$ plotted
	  as function of length scale for the CfA redshift survey. There
	  is no indication of a homogenization of the sample and both
	  $\Gamma(r)$ and $\xi(r)$ obey a power law, a result
	  consistent with a fractal structure for the distribution of
	  galaxies. The dashed line indicates a reference slope of $-1.7$
	  (Coleman, Pietronero and Sanders~1988).}
	  \label{2-15}
\end{figure}

Figure \ref{2-15} shows the calculation of $\Gamma(r)$,
$\Gamma^{\ast}(r)$ and $\xi(r)$ for the CfA redshift survey, where 
the absence of a homogenization of the distribution within the
sample and the absence of any kind of correlation length are clear.
This result brings strong support to the hypothesis that the
large scale distribution of galaxies forms indeed a fractal system. 

Table \ref{table-1-pmsl1996} shows the correlation properties of the galaxy 
distributions in terms of volume limited catalogues arising from most of the
$50,000$ redshift measurements that have been made to date. The samples are
statistically rather good in relation to the fractal dimension $D$ and the
conditional density $\Gamma(r)$, and their properties are in agreement with
each other.

\begin{figure}[p]
  \centerline{\epsffile{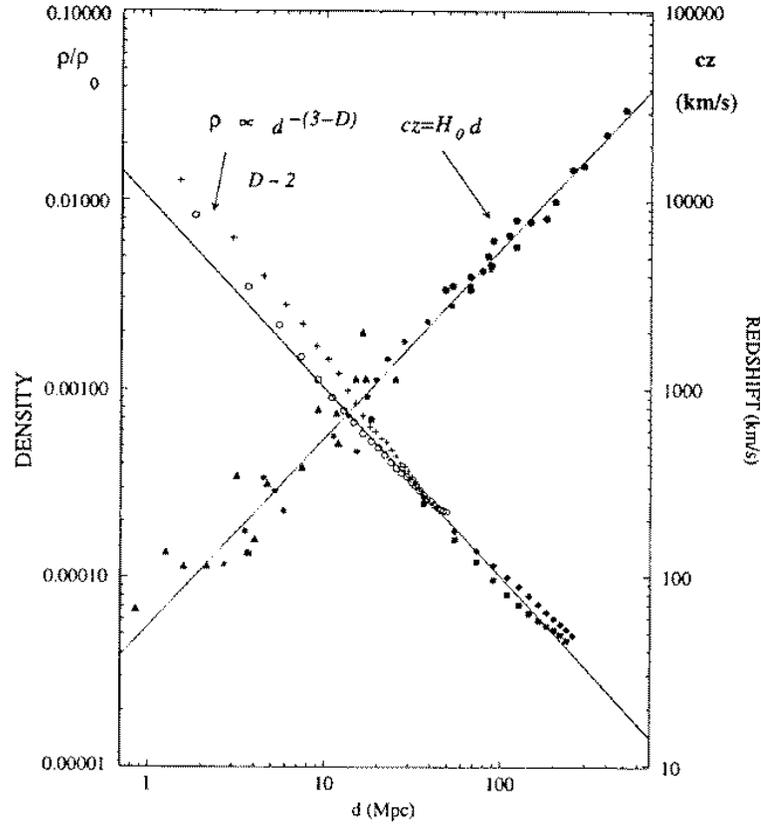}}
  \caption[\sf Graph of conditional average density and redshift versus
           distance] {\sf Conditional average density of galaxies plotted
           as function of distance (decreasing from left to right) for the
           following redshifts surveys: CfA1 (crosses), Perseus-Pisces
          (circles) and LEDA (squares). The solid line corresponds to
           $D = 2$. The Hubble redshift-distance is also shown in this
           graph. The dotted line corresponds to the Hubble law (increasing
           from left to right) with $H_0 = 55$ km s$^{-1}$ Mpc$^{-1}$. This
           law is constructed from: galaxies with Cepheid-distances for
           $cz > 0$ (triangles), galaxies with Tully-Fisher (B-magnitudes)
           distances (stars), galaxies with SNIa-distances for $cz > 3000$
           km/s (filled circles) (Pietronero, Montuori and Sylos Labini 1997).}
  \label{fig-4-pmsl1996}
\end{figure}

While various authors consider these catalogs as {\it not fair}, because
the contradiction between each other, Pietronero, Montuori and Sylos Labini 
(1997) show that this is due to the inappropriate methods of analysis. Figure 
\ref{fig-4-pmsl1996} shows a density power law decay for many redshift
surveys, and it is clear that we have well defined fractal 
correlations from $1$ to $1000$ h$^{-1}$ Mpc with fractal dimension
$D \approx 2$. This implies necessarily that the value of $r_0$ ($\xi (r_0)
= 1$) will scale with sample size $R_S$, \footnote{ \ See \S \ref{Power Law
Frac Dim} for an explanation of this effect under a fractal perspective.}
which gives the limit of the statistical validity of the sample, as
shown also from the specific data about $r_0$ in table \ref{table-1-pmsl1996}.
The smaller value of CfA1 was due to its limited size. At this same figure
we can see the so-called Hubble-de Vaucouleurs paradox, which is caused
by the coexistence between the Hubble law and the fractal distribution of
luminous matter at the same scales (Pietronero, Montuori and Sylos Labini
1997). \footnote{ \ See Ribeiro (1995) for a relativistic perspective of this
``paradox'', where it is shown that this is just a result of a relativistic 
effect.}

\begin{table}
 \caption{\sf The volume limited catalogues are characterized by the following
 parameters: $\Omega$ (sr) is the solid angle, $R_D$ (h$^{-1}$ Mpc) is the
 depth of the catalogue, $R_S$ (h$^{-1}$ Mpc) is the radius of the largest
 sphere that can be contained in the catalogue volume, $r_0$ (h$^{-1}$ Mpc)
 is the length at which $\xi(r_0) \equiv 1$, $D$ is the fractal dimension and
 $\lambda_0$ (h$^{-1}$ Mpc) is the eventual real crossover to homogeneity that
 this is actually never observed. The CfA2 and SSRS2 data are not yet
 available (Pietronero, Montuori and Sylos Labini 1997).}
 \label{table-1-pmsl1996}
 \begin{center}
 \begin{tabular}{|c||r|r|r|r|r|r|} \hline
  Sample  & $\Omega$~~ & $R_D$ & $R_S$ & $r_0$~ &     $D$~~~~   &   $\lambda_0$~~  \\ \hline
 \hline
 CfA1     &  1.83~     &  80   &  20   &  6~~   & $1.7 \pm 0.2$ & $ > 80$          \\ \hline
 CfA2     &  1.23~     & 130   &  30   & 10~~   &  2.0~~~~~~~   &    ?~~~          \\ \hline
 PP       &  0.9~~     & 130   &  30   & 10~~   & $2.0 \pm 0.1$ & $> 130$          \\ \hline
 SSRS2    &  1.13~     & 150   &  50   & 15~~   &  2.0~~~~~~~   &    ?~~~          \\ \hline
 LEDA     &  $4 \pi$~~ & 300   & 150   & 45~~   & $2.1 \pm 0.2$ & $> 150$          \\ \hline 
 LCRS     &  0.12~     & 500   &  18   &  6~~   & $1.8 \pm 0.2$ & $> 500$          \\ \hline       
 IRAS     &  $4 \pi$~~ &  80   &  40   &  4.5   & $2.0 \pm 0.1$ & $\sim 15$        \\ \hline
 ESP      &  0.006     & 700   &  10   &  5~~   & $1.9 \pm 0.2$ & $> 800$          \\ \hline
\end{tabular}
\end{center}
\end{table}

It is important to mention that there are effects which may conceal the
true statistical behaviour of the samples. Those effects may lead to the
conclusion that the sample under study is homogeneous, although such a
conclusion would be wrong, since such homogeneity may appear not as a
 statistical property of the sample, but just as an effect of its finite size.

\begin{figure}[p]
  \centerline{\epsffile{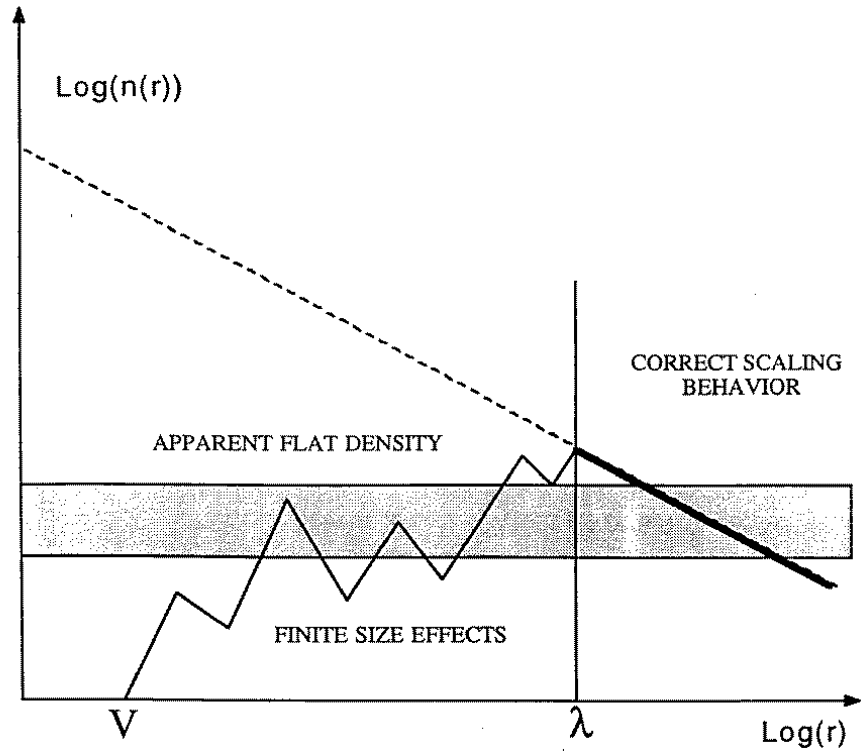}}
  \caption[\sf Schematic behaviour of the density computed from the vertex.]
      {\sf Schematic behaviour of the density computed from the vertex. Inside
     the Voronoi's length $V$ (small distances), one finds almost no galaxies.
     After this length the number of galaxies starts to grow with a regime
 strongly affected by finite size fluctuations, and the density can be
 approximately roughly by a constant value, leading to an apparent exponent
 $D \approx 3$. Finally the scaling region $r \geq \lambda$ is reached
 (Pietronero, Montuori and Sylos Labini 1997; Sylos Labini, Gabrielli,
 Montuori and Pietronero 1996).}
  \label{fig-6-pmsl1996}
\end{figure}

Figure \ref{fig-6-pmsl1996} shows the behaviour of the density computed 
from the vertex of a conic sample. At very small distances we are not going
to find any galaxy because the total number is rather small. After the
Voronoi's length $V$, which is of the order of the mean particle separation,
we begin to have signal but this is strongly affected by finite size effects.
The correct scaling behaviour is reached for the region $r \geq \lambda$.
In the intermediate region we have an apparent homogeneous distribution,
but it is due to the finite size effects (Pietronero, Montuori and Sylos
Labini 1997; Sylos Labini, Gabrielli, Montuori and Pietronero 1996; Sylos
Labini, Montuori and Pietronero 1998).

\begin{figure}[p]
  \centerline{\epsffile{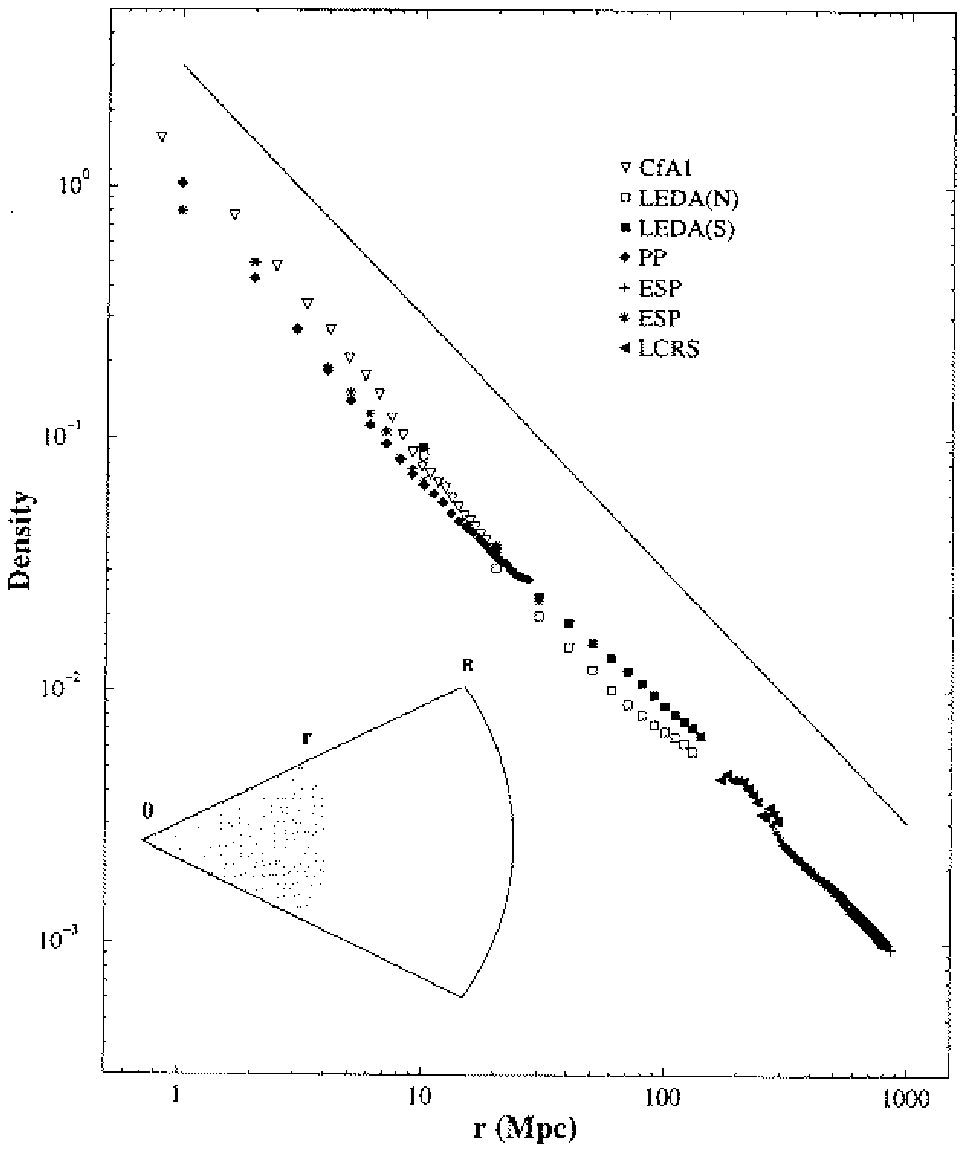}}
  \caption[\sf Graph of density versus distance showing full correlation for
 the various redshift catalogues] {\sf Full correlation for the various
 redshift catalogues in the range of distances $0.1 - 1 \times 10^3$
 h$^{-1}$ Mpc. A reference line with a slope $-1$ is also shown
 ($D \approx 2$). In the insert it is shown a conic volume. The radial
 density is computed by counting all the galaxies up to a certain limit,
 and by dividing for the volume of this conic sample (Pietronero, Montuori
 and Sylos Labini 1997).}
  \label{fig-7-pmsl1996}
\end{figure}

\begin{figure}[p]
  \centerline{\epsffile{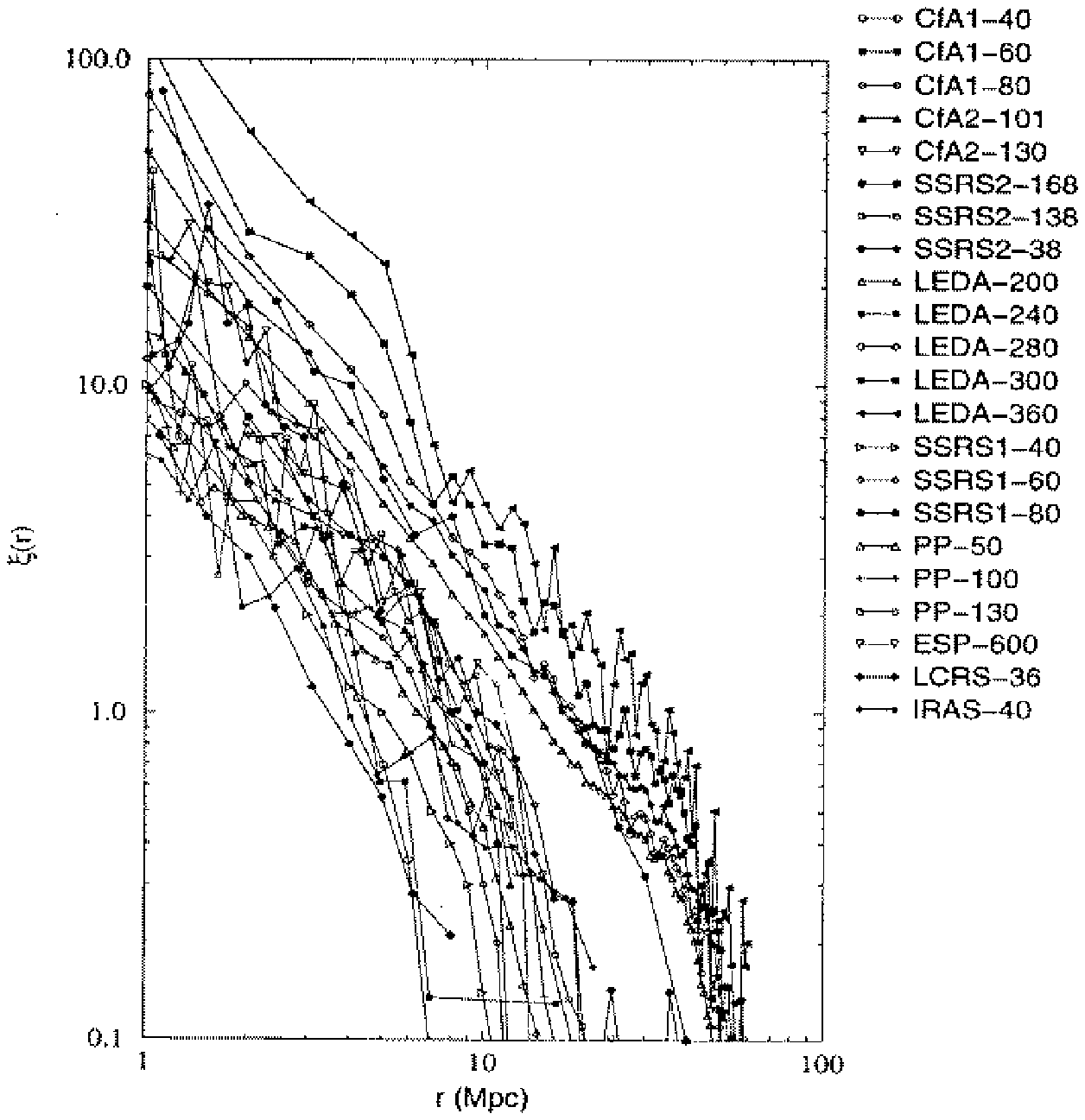}}
  \caption[\sf Graph of the function $\xi(r)$ versus the distance $r$ (Mpc).]
                    {\sf Graph of the function $\xi(r)$ versus the distance
 $r$  (Mpc) for the same galaxy catalogues of figure \ref{fig-7-pmsl1996}.
 Here we can see rather confusing results generated by the {\it a priori}
 and untested assumption of homogeneity, which are not present in the real
 galaxy distribution (Pietronero, Montuori and Sylos Labini 1997).} 
  \label{fig-8-pmsl1996}
\end{figure}

As evidence that this new statistical approach is the appropriate method of 
analysis, we can see in figure \ref{fig-7-pmsl1996} an agreement between 
various available redshift catalogues in the range of distances
$0.1 - 1 \times 10^3$ h$^{-1}$ Mpc. From this we can conclude that there
is no tendency to homogeneity at this scale. In contrast to figure
\ref{fig-7-pmsl1996} we have figure \ref{fig-8-pmsl1996} where the
traditional analysis based on $\xi(r)$ for the same catalogues of figure
\ref{fig-7-pmsl1996} shows a strong conflict between these two analytical
methods. Figure \ref{fig-8-pmsl1996} shows that $\xi(r)$ is unable to
say without any doubt if there is a homogeneous scale, and this is the
main reason from where the concept of {\it ``unfair sample''} is generated.

\begin{figure}[p]
  \centerline{\epsffile{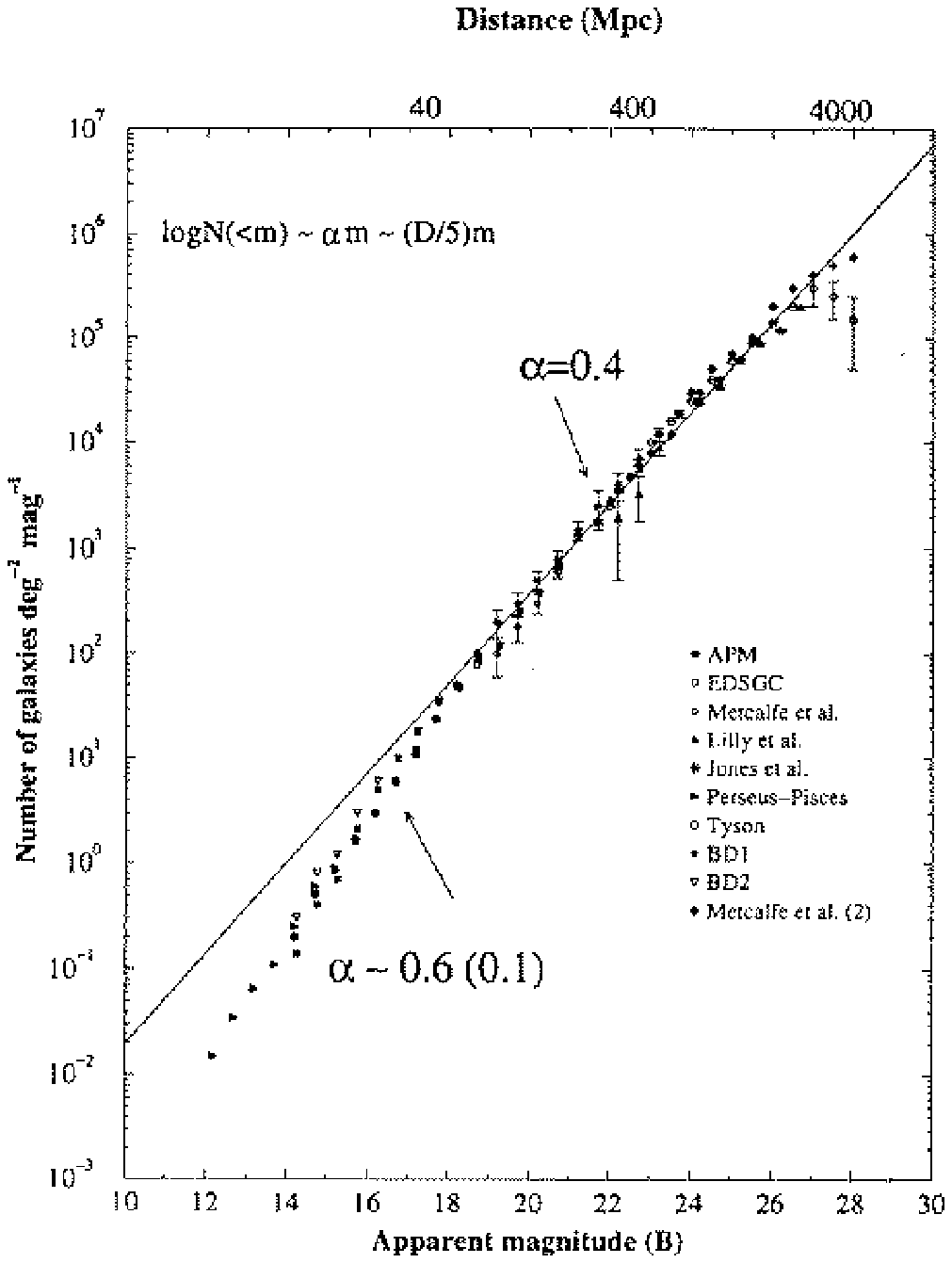}}
  \caption[\sf Graph of the galaxy number counts in the {\it B}-band from
 several surveys versus the apparent magnitude.] {\sf The galaxy number
 counts in the {\it B}-band from several surveys. In the range $12 \leq m
 \leq 19$ the counts show an exponent $\alpha = 0.6 \pm 0.1$, while in the
 range $19 \leq m \leq 28$ the exponent is $\alpha \approx 0.4$. The
 amplitude of galaxy number counts for $m \geq 19$ (solid line) is computed
 from the determination of the prefactor {\it B} of the density $n(r) = B
 r^{-(3-D)}$ at small scale and from the knowledge of the galaxy luminosity
 function. The distance is computed for a galaxy with $M = -16$ and the
 value used for $H_0$ is $75$ km s$^{-1}$ Mpc$^{-1}$ (Pietronero, Montuori
 and Sylos Labini 1997).}
 \label{fig-9-pmsl1996}
\end{figure}

Another statistical analysis is the number counts and angular correlations. 
In figure \ref{fig-9-pmsl1996} we show the behaviour of the number counts 
versus magnitude relation ($N (< m)$) with an exponent 
$\alpha = D / 5$. At small scales $\alpha = 0.6 \pm 0.1$ ($D \approx 3$), 
which means that we have an apparent homogeneity. However, this is due to the 
finite size effects discussed above, while at larger scales the value $0.4$ ($D 
\approx 2$) shows correlation properties of the sample in agreement with 
the results obtained for $\Gamma(r)$. In addition, the fact that the
exponent $0.4$ holds up to magnitudes $27 - 28$ seems to indicate that
the fractal structure may continue up to $2 - 3 \times 10^3$ h$^{-1}$ Mpc
(Pietronero, Montuori and Sylos Labini 1997).

\subsection{Pietronero-Wertz's Single Fractal (Hierarchical) 
 Model} \label{Pietr-Wertz's Model}

The single fractal model proposed by Pietronero (1987) is essentially an
application of the cluster fractal dimension to the large scale
distribution of galaxies, a straightforward analysis of the consequences
of this fractal interpretation of the galactic system plus the proposal
of new statistical tools to analyse
the catalogues of galaxies, derived from his strong criticisms of the
standard statistical analysis based on the two-point correlation
function. He also studied the problem from a multifractal perspective. 
What we shall attempt to do here is to present a summary of the basic 
results related to the discussion above. Later in a review paper, Coleman 
and Pietronero (1992) extended the theory, with especial emphasis 
on multifractals and the angular correlation function, and added new results. 
Further extensions and results of this theory were also made in Sylos
Labini, Montuori and Pietronero (1998).

Wertz's (1970, 1971) hierarchical model, on the other hand, was proposed
at a time when fractal ideas had not yet appeared. However, these ideas
were more or less implicit in his work, and as we shall see below, he
ended up proposing a hierarchical model mathematically identical to
Pietronero's single fractal model, but 17 years earlier. For this reason
his work deserves to be quoted as being the first independent model where
self-similar ideas were applied in the study of the large scale distribution
of galaxies. The reasons why Wertz's work was forgotten for so long lies
on the shortcomings of its physical implications, as it will be shown below.

\subsubsection{Pietronero's Single Fractal Model}\label{Single Frac Model}

\begin{figure}[htb]
  \centerline{\epsffile{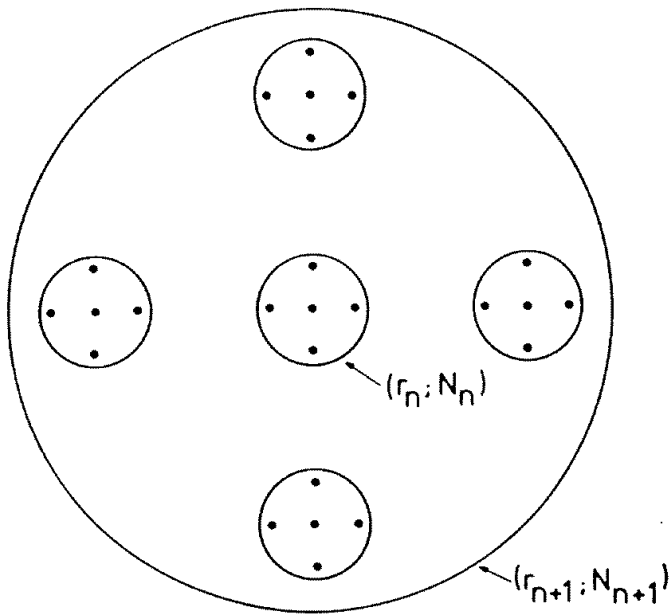}}
  \caption[\sf Schematic illustration of a deterministic fractal system
          from where a fractal dimension can be derived (Pietronero
	  1987).]{\sf Schematic illustration of a deterministic
	  fractal system from where a fractal dimension can be derived.
	  The structure is self-similar, repeating itself at different
	  scales (Pietronero 1987).}
	  \label{2-13}
\end{figure}

The definition of the fractal dimension as made by Pietronero (1987) is
slightly different than in the case of the cluster dimension discussed
previously. Figure \ref{2-13} shows a schematic
representation of a fractal distribution of points. Clearly we have a
self-similar distribution as more and more structure appears at smaller
and smaller scales, with the structure at small scales being similar to
the one at large scales. By starting from a point occupied by an object
and counting how many objects are present within a volume characterized
by a certain length scale, we have that within a certain radius $d_0$,
there are $N_0$ objects; then within $d_1~=~k~d_0$ there are
$N_1~=~\tilde{k}~N_0$ objects; in general, within
\begin{equation}
   d_n = k^n d_0
\label{e2}
\end{equation}  
we have
\begin{equation}
   N_n = \tilde{k}^n N_0
    \label{e1}
\end{equation}
objects. Generalizing this idea to a smooth relation, we can define a
{\it generalized~mass-length~relation} between $N$ and $d$ of the type
\begin{equation}
   N(d) = \sigma d^D,
   \label{eq-ch2-11}
\end{equation}
where the fractal dimension
\begin{equation}
  D = \frac{\log \tilde{k}}{\log k}
  \label{eq-ch2-12}
\end{equation}
depends only on the rescaling factors $k$ and $\tilde{k}$, and the
prefactor $\sigma$ is related to the lower cutoffs $N_0$ and $d_0$,
\begin{equation}
  \sigma = \frac{ N_0}{ {d_0}^D}.
   \label{eq-ch2-13}
\end{equation}
Equation (\ref{eq-ch2-11}) corresponds to a continuum limit for the
discrete scaling relations.

Let us now suppose that a sample of radius $R_S$ contains a portion of
the fractal structure. If we assume it to be a sphere, then the sample
volume is given by 
\[
 V(R_S) = \frac{4}{3} \pi {R_S}^3,
\]
which allows us, together with equation (\ref{eq-ch2-11}), to compute
the average density $\langle n \rangle$ as being
\begin{equation}
  \langle n \rangle = \frac{N (R_S)}{V (R_S)} =
		      \frac{3 \sigma}{4 \pi} {R_S}^{- \gamma}, \ \ \ \
                      \gamma = 3 - D.
  \label{eq-ch2-14}
\end{equation}
This is the same type of power law expression obtained several years ago
by de Vaucouleurs (1970), and equation (\ref{eq-ch2-14}) shows very
clearly that the average density is not a well defined physical property
for this sort of fractal system because it is a function of the sample
size.

\subsubsection{Wertz's Hierarchical Model}\label{Wertz's Model}

\begin{figure}[p]
  \centerline{\epsffile{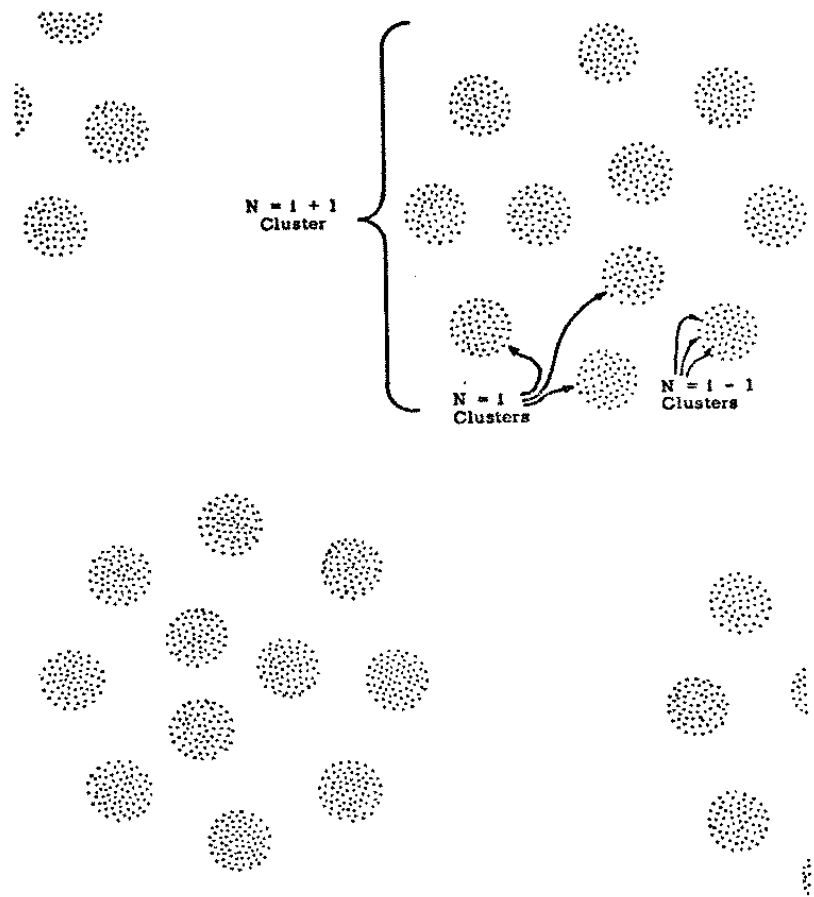}}
  \caption{\sf Reproduction from Wertz (1970, p.\ 25) of a rough
	   sketch cross-section of a portion of an $N=i+2$ cluster of a
	   polka dot model.}\label{figura-wertz}
\end{figure}

The hierarchical model advanced by Wertz (1970, 1971) was conceived at a
time when fractal ideas had not yet appeared. So, in developing his
model, Wertz was forced to start with a more conceptual discussion in order to
offer ``a clarification of what is meant by the `undefined notions' which
are the basis of any theory'' (Wertz 1970, p.\ 3). Then he stated that
``a cluster consists of an aggregate or gathering of {\sc elements} into a
more or less well-defined group which can be to some extent distinguished
from its surroundings by its greater density of elements. A {\sc hierarchical}
structure exists when {\sc {\rm i}th order clusters} are themselves elements
of an {\sc {\rm (i+1)}th order cluster}. Thus, galaxies ({\sc zeroth order
clusters}) are grouped into {\sc first order cluster}. First order clusters
are themselves grouped together to form {\sc second order clusters}, etc,
{\it ad infinitum}'' (see figure \ref{figura-wertz}).

Although this sort of discussion may be very well to start with, it
demands a precise definition of what one means by a cluster in order to
put those ideas on a more solid footing, otherwise the hierarchical
structure one is talking
about continues to be a somewhat vague notion. Wertz seemed to have realized
this difficulty when later he added that ``to say what percentage
of galaxies occur in clusters is beyond the abilities of current
observations and involves the rather arbitrary judgment of what sort of
grouping is to be called a cluster. (...) It should be pointed out that
there is not a clear delineation between clusters and superclusters''~(p.\ 8).

Despite this initially descriptive and somewhat vague discussion about
hierarchical structure, which is basically a discussion about scaling
in the fractal sense, Wertz did develop some more precise notions when
he began to discuss specific models for hierarchy, and his starting
point was to assume what he called the ``universal density-radius
relation'', that is, the de Vaucouleurs density power law, as a
fundamental empirical fact to be taken into account in order to develop
a hierarchical cosmology. Then if $M(x,r)$ is the total mass within a
sphere of radius $r$ centered on the point $x$, he defined the {\it volume
density} $\rho_v$ as being the average over a sphere of a given volume
containing $M$. Thus
\begin{equation}
 \rho_v (x,r) \equiv \frac{3 M(x,r)}{4 \pi r^3},
 \label{e6}
\end{equation}
and the {\it global density} was defined as being 
\begin{equation}
 \rho_g \equiv \lim_{r \rightarrow \infty} \rho_v (x,r).
 \label{e7}
\end{equation}
A {\it pure hierarchy} is defined 
as a model universe which meets the following pos\-tu\-lates:

\noindent {\it (i)}
for any positive value of $r$ in a bounded region, the volume density
has a maximum;

\noindent  {\it (ii)} the model is composed of only
mass points with finite non-zero mean mass;

\noindent  {\it (iii)}
the zero global density postulate: ``for a pure hierarchy the global
density exists and is zero everywhere'' (see Wertz 1970, p.\ 18).

With this picture in mind, Wertz states that ``in any model which involves
clustering, there may or may not appear discrete lengths which represent
clustering on different scales. If no such scales exist, one would have an
{\sc indefinite hierarchy} in which clusters of every size were equally
represented (...). At the other extreme is the {\sc discrete hierarchy} in
which cluster sizes form a discrete spectrum and the elements of one size
cluster are all clusters of the next lowest size'' (p.\ 23). Then in order 
to describe {\it polka dot models}, that is, structures in a discrete hierarchy
where the elements of a cluster are all of the same mass and are distributed
regularly in the sense of crystal lattice points, it becomes necessary for
one be able to assign some average properties. So if $N$ is the order of
a cluster, $N=i$ is a cluster of arbitrary order (figure~\ref{figura-wertz}),
and at least in terms of averages a cluster of mass $M_i$, diameter $D_i$ and
composed of $n_i$ elements, each of mass $m_i$ and diameter $d_i$, has
a density given by
\begin{equation}
 \rho_i = \frac{6 M_i}{\pi {D_i}^3}.
 \label{e8}
\end{equation}
From the definitions of discrete hierarchy it is obvious that
\begin{equation}
 M_{i-1} = n_{i-1} m_{i-1} = m_i,
 \label{e9}
\end{equation}
and if the ratio of radii of clusters is
\begin{equation}
 a_i \equiv \frac{D_i}{d_i} = \frac{D_i}{D_{i-1}},
 \label{e10}
\end{equation}
then the {\it dilution factor} is defined as 
\begin{equation}
 \phi_i \equiv \frac{\rho_{i-1}}{\rho_i} = \frac{{a_i}^3}{n_i} > 1,
 \label{e11}
\end{equation}
and the {\it thinning rate} is given by
\begin{equation}
 \theta_i \equiv \frac{\log (\rho_{i-1} / \rho_i)}{\log (D_i / D_{i-1})}
          =  \frac{ \log ({a_i}^3 / {n_i} ) }{\log a_i}.
 \label{e12}
\end{equation}

A {\it regular polka dot model} is defined as the one whose number of
elements per cluster $n_i$ and the ratio of the radii of successive clusters
$a_i$ are both constants and independent of $i$, that is, $n$ and $a$
respectively. Consequently, the dilution factor and the thinning rate are
both constants in those models,
\begin{equation}
  \phi=\frac{a^3}{n}, \ \ \ \ \theta=\frac{\log (a^3 /n)}{\log a}.
 \label{e13}
\end{equation}

The {\it continuous representation} of the regular polka dot model, which
a\-mounts essentially to writing the hierarchical model as a continuous
distribution, is obtained if we consider $r$, the radius of spheres
centered on the origin, as a continuous variable. Then, from equation
(\ref{e10}) the radius of the elementary point mass $r_0$, is given by 
\begin{equation}
 r_0 = \frac{R_1}{a},
 \label{e14}
\end{equation}
where $R_N$ is the radius of a Nth order cluster with $M_N$ mass,
$V_N$ volume, and obviously that $R_0=r_0$. It follows from equation
(\ref{e14}) the relationship between $N$ and $r$,
\begin{equation}
 r= a^N r_0,
 \label{e15}
\end{equation}
where $R_N=r$. 

Notice that by doing this continuous representation Wertz 
ended up obtaining an equation (eq.\ \ref{e15}) which is nothing more
than exactly equation (\ref{e2}) of Pietronero's single fractal model,
although Wertz had reached it by means of a more convoluted reasoning.
Actually, the critical hypothesis which makes his polka dot model
essentially the same as Pietronero's fractal model was the assumption of
regularity of the model because, in this case $a$ and $n$ become constants. Also
notice that this continuous representation amounts to changing from 
discrete to an indefinite hierarchy, where in the latter the characteristic
length scales for clustering are absent. Therefore, in this
representation clusters (and voids) extend to all ranges where the
hierarchy is defined, with their sizes extending to all scales between
the inner and possible outer limits of the hierarchy. Hence, in this
sense the continuous representation of the regular polka dot model has
exactly the same sort of properties as the fractal model discussed by
Pietronero.

From equation (\ref{e9}) we clearly get
\begin{equation}
 M_N = n^N M_0,
 \label{e16}
\end{equation}
which is equal to equation (\ref{e1}), except for the different notation,
and hence the de Vaucouleurs density power law is easily obtained as
\begin{equation}
 \rho_v = \frac{M_N}{V_N} = \left[ \frac{3 M_0}{4 \pi {r_0}^{ ( \log n /
           \log a ) }} \right] r^{- \theta },
 \label{e17}
\end{equation}
where $\theta$ is the thinning rate
\begin{equation}
   \theta =  3 - \left( \frac{ \log n}{ \log a} \right).
   \label{e18}
\end{equation}
Notice that equations (\ref{e17}) and (\ref{e18}) are exactly equations
(\ref{eq-ch2-14}), where $\gamma$ is now called the thinning rate. Finally, the
{\it differential density}, called {\it conditional density} by Pietronero, 
is defined as
\begin{equation}
 \rho_d \equiv \frac{1}{4 \pi r^2} \frac{ d M(r)}{d r} = 
	\left( 1 - \frac{\theta}{3} \right) \rho_v.
 \label{e19}
\end{equation}

From the presentation above it is then clear that from a geometrical
viewpoint Wertz's continuous representation of the regular polka dot
model is nothing more than Pietronero's single fractal model. However,
the two approaches may be distinguished from each other by some
important conceptual differences. Basically, as Pietronero clearly
defines the exponent of equation (\ref{eq-ch2-11}) as a fractal dimension,
that immediately links his model to the theory of critical phenomena in
physics, and also to nonlinear dynamical systems, bringing a completely
new perspective to the study of the distribution of galaxies, with
potentially new mathematical concepts and analytical tools to investigate
this problem. In addition, he strongly emphasized the fundamental
importance of scaling behaviour in the observed distribution of galaxies
and the fundamental role of the exponent of the power law, as well as
pointing out the appropriate mathematical tool to describe this
distribution, namely the fractal dimension. Finally, as many fractals
have a statistical nature, either in their description or in their
construction, or both, and also are inhomogeneously distributed, the
statistical methods capable of dealing with fractals must also be able
to derive well-defined statistical properties even from inhomogeneous
samples, fractals or not, where the average density may not be
well-defined. Therefore, the fractal perspective
brings together a completely new set of statistical tools capable of a
comprehensive reinterpretation of the conclusions drawn upon the
available data about the distribution of galaxies, and without any
need of {\it a priori} assumptions about homogeneity.

All that is missing in Wertz's approach, and his thinning
rate is just another parameter in his description of hierarchy, without
any special physical meaning attached to it. Therefore, in this sense
his contribution started and remained as an isolated work, ignored
by most, and which could even be viewed simply as an ingenious way
of modelling Charlier's hierarchy, but nothing more.

Nonetheless, it should be said that this discussion must not be
viewed as a critique of Wertz's work, but simply as a realization of the
fact that at Wertz's time nonlinear dynamics and fractal geometry were
not as developed as at Pietronero's time, if developed at all, and
therefore Wertz could not have benefitted from those ideas. Despite this
it is interesting to note that even with less data and mathematical concepts
he was nevertheless able to go pretty far in discussing scaling behaviour
in the distribution of galaxies, developing a model to describe it in the
context of Newtonian cosmology, and even suggesting some possible ways of
investigating relativistic hierarchical cosmology.

\subsection{Consequences of the Single Fractal Model}

As stated above, the fractal model shown in the previous section offers an
attractively simple explanation for the results obtained when analysing
the distribution of galaxies by means of the new statistical methods
advanced by Pietronero and collaborators. Thus, when this statistic is
applied to a fractal system some important results that can be related
to the observed distribution of galaxies are obtained. Here in this
subsection we shall discuss some straightforward results arising from
the fractal approach to the galaxy clustering problem. But, before we
start this discussion some important remarks must be made.

Firstly, although the fractal distribution never becomes homogeneous, it is 
a {\it statistically fair sample} in the sense mentioned above, which is 
contrary to the traditional lines where only a homogeneous sample is
taken to be a fair one. 

Secondly, a three dimensional galaxy distribution (figures
\ref{fig-3a,3b,3c-pmsl1996}a, b), which has fractal properties when
studied in 3D, appears relatively homogeneous at some large angular scale
(figure \ref{fig-3a,3b,3c-pmsl1996}c), loosing some of its irregular
characteristics when projected on an angular distribution. Due to this
property of fractal structures, it is necessary great care when dealing
with projected structures, as their fractal features may become hidden when
dealing with 2D data (Coleman and Pietronero 1992; Pietronero,
Montuori and Sylos Labini 1997).

Thirdly, the galaxy distribution has been studied also in terms of their mass  
(Pietronero 1987; Coleman and Pietronero 1992) and their luminosity
distribution (Pietronero, Montuori and Sylos Labini 1997), which is a
full distribution and not a simple set of points. The study of these
distributions requires a generalization of the fractal dimension and the
use of the concept of multifractals. In this case, we have different
scaling properties for different regions of the system, while in the
fractal case only one exponent characterizes the entire system. A
multifractal analysis also shows a new important relation between the
luminosity function (Schechter luminosity distribution) and the space
correlation properties.

\begin{figure}[p]
  \centerline{\epsffile{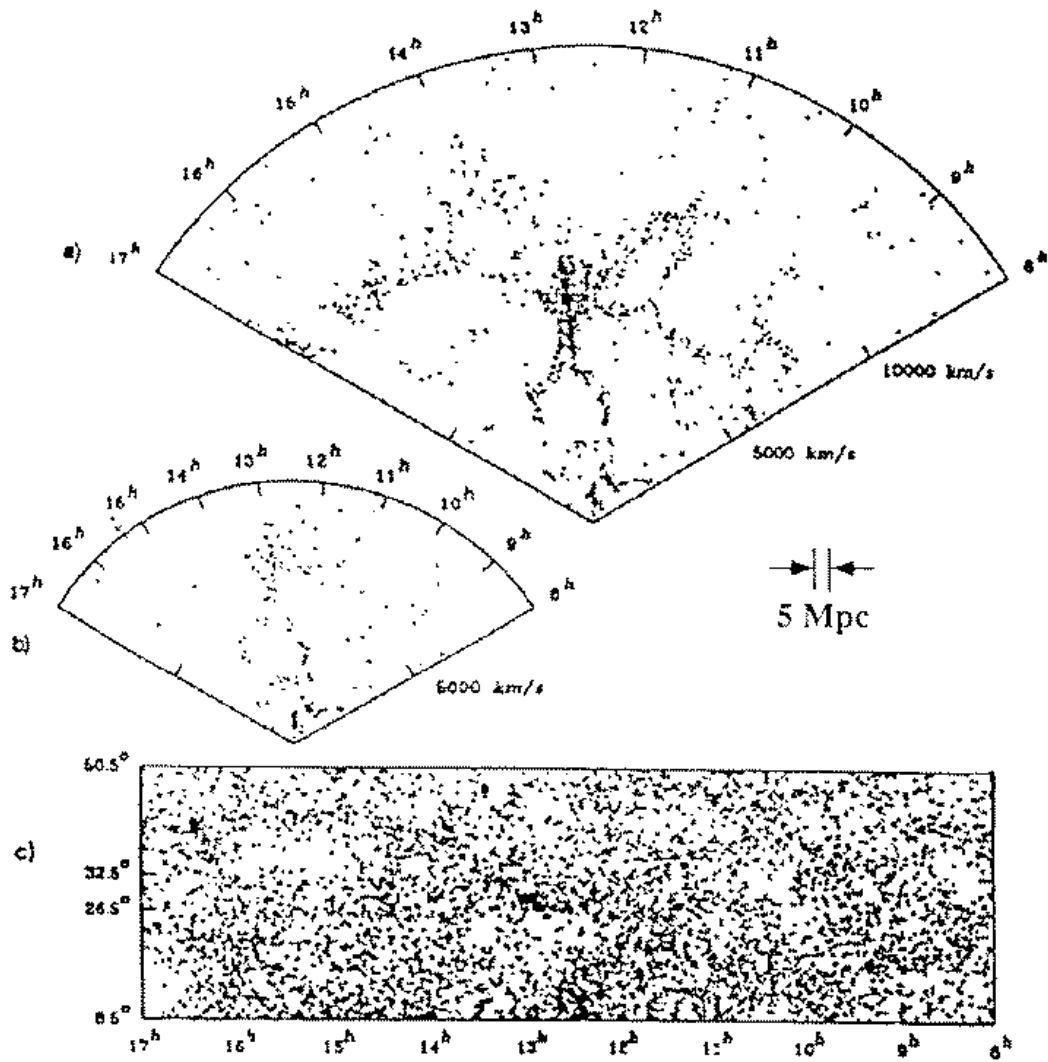}}
  \caption[\sf Comparison between the angular distribution and the three
  dimensional galaxy distribution (Pietronero, Montuori and Sylos Labini
 1996).]{\sf A slice of the three dimensional galaxy distribution (a) and (b)
 compared with the corresponding (c) angular distribution. Note that the
 angular distribution appears relatively homogeneous while the real three
 dimensional distribution in space is much more irregular. This picture
 shows the Great Wall which extends over the entire sample (at least
 $170$ h$^{-1}$ Mpc) (Pietronero, Montuori and Sylos Labini 1997).}    
  \label{fig-3a,3b,3c-pmsl1996} 
\end{figure}

Finally,
equation (\ref{eq-ch2-14}) shows
the intrinsic inhomogeneity of the fractal system, and although we cannot
yet rule out the possibility of an upper cutoff to homogeneity, if the fractal
system were unlimited the average density will tend to zero at increasing
distances. This possibility has provoked strong reactions from some authors
as they assumed it to go against established ideas in cosmology and,
therefore, could not accept it. Nevertheless, it is important to point out
that such a traditional view is not as sound as it seems at first, inasmuch
as Ribeiro (1992, 1993, 1994) showed that even Friedmannian cosmologies
do allow Wertz's zero global density postulate under a specific
relativistic interpretation.

\subsubsection{Power Law Behaviour and Fractal Dimension}\label{Power 
Law Frac Dim}

Pietronero (1987) defined the conditional density from an occupied point
of the fractal system as being given by
\begin{equation}
 \Gamma (r) = \frac{1}{S(r)} \frac{dN(r)}{dr} = \frac{D \sigma}{4 \pi}
              r^{- \gamma},
  \label{eq-ch2-15}
\end{equation}
where $S(r)$ is the area of the spherical shell of radius $r$ (this
is the same as Wertz's differential density). Now by
means of equations (\ref{eq-ch2-2-11b}) and (\ref{eq-ch2-14}), it
is straightforward to see that
\begin{equation}
 \xi (r) = \left( \frac{3 - \gamma}{3} \right)
	   { \left( \frac{r}{R_S} \right) }^{- \gamma} -1.
 \label{eq-ch2-16}
\end{equation}
The two equations above show clearly the dependence of $\xi(r)$ on the
sample size $R_S$ while this is not the case for $\Gamma(r)$. It is
clear therefore, that in a fractal system the function $\xi(r)$ mixes
up the physical properties of the distribution with the sample size,
an effect actually observed in analysis of catalogues, as seen above.

Since the correlation length $r_0$ is defined as the point at which
$\xi(r)~=~1$, we have
\begin{equation} 
  r_0 = { \left( \frac{3 - \gamma}{6} \right) }^{1/\gamma} R_S,
 \label{eq-ch2-17}
\end{equation}
which is dependent on $R_S$. Therefore, the fractal scenario
explains the dependence of the correlation length $r_0$ with the sample
size. The point at which $\xi(r)~=~0$ is
\begin{equation}
 {r'}_0 = { \left( \frac{3 - \gamma}{3} \right) }^{1/\gamma} R_S.
  \label{eq-ch2-18}
\end{equation}
For $r~>~{r'}_0$ the function $\xi(r)$ is negative and for $r~\ll~R_S$
it can be approximated by a power law,
\begin{equation} 
 \xi (r) \approx A(R_S) r^{- \gamma}, \ \ \ \ \ (r \ll r_0),
 \label{eq-ch2-19}
\end{equation}
where the two-point correlation function amplitude is given by 
\begin{equation}
 A(R_S) = { \left( \frac{3 - \gamma}{3} \right) } {R_S}^\gamma.
 \label{eq-ch2-20}
\end{equation}
Considering equation (\ref{eq-ch2-10}) obtained empirically, we then
have a fractal dimension $D~\approx~1.3$ for the distribution of
galaxies in this model for small $r$. For larger $r$ the fractal dimension
is $D~\approx~2$.

From these simple equations it is clear that with the exception of the
exponent $\gamma$, all relevant features of the function $\xi(r)$ are
related to the sample radius $R_S$ and not to the intrinsic properties
of the system.

Equation (\ref{eq-ch2-20}) allows us to obtain a new interpretation of
the difference in the amplitude of the correlation (\ref{eq-ch2-19}).
Considering now a sample radius $R_G$ for the galaxy catalogue, the
amplitude of the two-point correlation function is clearly given by 
\[
 A_G = { \left( \frac{3 - \gamma}{3} \right) } {R_G}^\gamma.
\]
In the case of clusters we then have
\[ 
 A_C = { \left( \frac{3 - \gamma}{3} \right) } {R_C}^\gamma.
\]

The hypothesis of self-similarity means that it is the exponent of the
power law  which matters as giving the property of the structure. The
amplitude is just a rescaling factor related to the size of the sample
and the lower cutoff, without a direct physical meaning. As an example,
under a rescaling of the length by a factor $b$ such that
$r~\to~r'=b~\cdot~r$, a self-similar function will be
rescaled as the functional relation
$f(r')=f(b~\cdot~r)~\equiv~A(b)~\cdot~f(r)$. This is clearly satisfied
by power laws with any exponent. In fact, for $f(r)=f_0r^D$ we have
$f(r')=f_0{(br)}^D={(b)}^D~\cdot~f(r)$. Therefore, under the assumption
of a single self-similar structure, the amplitudes $A_G$ and $A_C$
should be related by
\begin{equation}
 \frac{A_C}{A_G} = { \left( \frac{R_C}{R_G} \right) }^{\gamma}.
 \label{eq-ch2-21}
\end{equation}
Since for the galaxy and cluster catalogues we have $R_C~\approx~5R_G$,
equation (\ref{eq-ch2-21}) predicts (with $\gamma=1.8$)
\[
 \frac{A_C}{A_G} \approx 18,
\]
which is the value found for the discrepancy in the amplitudes. Note
that this is a simple evaluation of the mismatch by a single
deterministic fractal which does not take into account any stochastic
process, which would be a more realistic situation in the case of the
distribution of galaxies. This discussion shows therefore that
correlations of clusters appear to be a continuation of galaxy
correlations in larger scales, and the discrepancy in amplitudes simply
means different observations of the same system at different depth
samples.

So we see that the fractal hypothesis explains many of the
puzzling problems so far encountered in the study of the large scale 
distribution of galaxies. It is important to say that this is
accomplished with a model of remarkable simplicity, showing once more the
strength of the fractal concept of dealing with this sort of complex
problems. Therefore, from a theoretical point of view,
due to the simplicity of the model and the strength of the concept, it
becomes seductive to apply a fractal approach to modern cosmology.

\subsubsection{Possible Crossover to Homogeneity}\label{Cross Homo}

Until now we have seen that this new statistical approach shows no evidence
for a homogeneous distribution of the visible matter on large scale, but we
cannot yet exclude this possibility, as it may occur at some scale not yet
observed.

As we saw in \S \ref{Corr Anal without Assump} 
the conditional density $\Gamma (r) = G(r) / \langle n \rangle$, 
obeys a power law for $r<\lambda_0$, and for $r\geq\lambda_0$ it is constant.

By equation (\ref{eq-ch2-15}), we have
\begin{equation}
 \Gamma (r) = \frac{D \sigma}{4 \pi} r^{- (3-D)}, \hspace{5mm} r<\lambda_0,
  \label{eq-5.21-cp1992}
\end {equation}
\begin{equation}
 \Gamma (r) = n_0,\hspace{5mm} r\geq\lambda_0. 
 \label{eq-5.22-cp1992}
\end{equation} 
Then at $\lambda_0$, we have
\begin{equation}
 n_0 = \frac{D \sigma}{4 \pi}
              {\lambda_0}^{- (3-D)}. 
 \label{eq5.23-cp1992}
\end{equation}
Therefore the function $\Gamma (r)$ has a power law behaviour up to 
$\lambda_0$ and it is constant thereafter.

The integrated conditional density (\ref{eq-ch2-2-11d}) may be written as
\begin{equation}
 I(r) = \sigma r^D,\hspace{5mm} r<\lambda_0,
 \label{eq-5.24-cp1992}
\end{equation}
\begin{equation}
 I(r) = D \sigma {\lambda_0}^D \left(\frac{1}{D}-\frac{1}{3}\right) + \frac{1}{3} 
D \sigma {\lambda_0}^{- (3-D)}, \hspace{5mm} r\geq\lambda_0,         
 \label{eq5.25-cp1992}  
\end{equation}
which is the total number of galaxies. The change in the power law from $D$ to 
$3$ indicates the crossover to length $\lambda_0$.

If we want to consider the function $\xi(r)$ for a system with a crossover to
homogeneity we have to specify our sample radius $R_S$ ($R_S > \lambda_0)$ 
explicitly. Considering that the density $\langle n \rangle$ is the total number of 
galaxies per volume unity of the sample, we have
\begin{equation}
 \langle n \rangle = \frac{D \sigma}{4 \pi}{\lambda_0}^{- (3-D)}
\left[1+\left(\frac{3}{D}-1\right){\left(\frac{\lambda_0}{R_S}\right)}^3\right]
 \label{eq-5.26-cp1992}
\end{equation}
and, considering equation (\ref{eq-5.21-cp1992}) we obtain
\begin{equation}
 \xi(r) = \frac{D \sigma}{4 \pi} \frac{r^{- (3-D)}}{\langle n \rangle} - 1, 
\hspace{5mm} r<\lambda_0,
 \label{eq-5.27-cp1992}
\end{equation}
\begin{equation}
 \xi(r) = \frac{D \sigma}{4 \pi} \frac{{\lambda_0}^{- (3-D)}}{\langle n \rangle} - 
1, \hspace{5mm} 
 r \geq\lambda_0,
 \label{eq-5.28-cp1992}
\end{equation} 
where the dependence of $\langle n \rangle$ on $R_S$ implies that the point at 
which $\xi(r) = 0$ is a function of $\lambda_0$ and $R_S$. Then, the function 
$\xi(r)$, which is still inappropriate, becomes an appropriate tool only in the 
limiting case where $\lambda_0 \ll R_S$, in the sense that $\langle n \rangle$ 
becomes independent of $R_S$. Then, in this case we have
\begin{equation}
 \langle n \rangle =  \frac{D \sigma}{4 \pi} {\lambda_0}^{- (3-D)},
 \label{eq-<n>}
\end{equation}
\begin{equation}
 \xi(r) = {\left(\frac{r}{\lambda_0}\right)}^{-(3-D)} - 1,\hspace{5mm}  
r<\lambda_0,
\end{equation}
\begin{equation}
 \xi(r) = 0,\hspace{5mm}  r\geq \lambda_0.
\end{equation}
Only in this case the length $r_0$ is related to the correlation length 
$\lambda_0$, 
\begin{equation}
 r_0 = 2^{1 / (D-3)} \lambda_0.
\end{equation}
For the case $D\approx 2$ one has $r_0 = {\lambda_0 / 2}$. 

\section{Conclusions}

In this paper we have studied the distribution of galaxies by means of
a Newtonian fractal perspective. We began with a brief introduction to fractals,
and in \S \ref{Distr Galaxies} we discussed a more appropriated
statistical analysis for the large scale distribution of galaxies. We
also discussed the fractal hypothesis itself which contrasts to the
orthodox traditional view of an {\it observationally} homogeneous universe.
We finished this section showing the Pietronero-Wertz's single fractal
(hierarchical) model for describing and analysing the large scale
distribution of galaxies and some of its consequences. In summary the
main results of \S \ref{Distr Galaxies} are:
\begin{enumerate}
\item the employment of the two-point correlation function $\xi(r)$ for a
      statistical analysis of the galaxy distribution is problematic due
      to the {\it a priori} assumption of homogeneity. This assumption is
      hidden in the average density which does not seem to be a well
      defined physical property for the galaxy distribution because it is
      a function of the sample size, a result consistent with the fractal
      view of such a distribution. In order to solve this problem
      Pietronero (1987) introduced a new statistical approach for the
      galaxy distribution, which can test the assumption of homogeneity;
\item the comparison between two different methods of analysis shows that
      an inappropriate method leads to wrong conclusions in volume
      limited catalogues, due to its finite size and shallowness. Due to
      this they may wrongly be considered as not fair, as shown in
      table \ref{table-1-pmsl1996}. So, an apparent homogeneous distribution
      in the region $V \leq r < \lambda$ of figure \ref{fig-6-pmsl1996},
      where $V$ is the Voronoi's length and $\lambda$ is the inferior
      limit for reaching the correct scaling behaviour, is due to the
      finite size effects, and the correct scaling is reached for the
      region $r \geq \lambda$;
\item homogeneity must be regarded as a property of the sample and not a
      condition of its statistical validity;
\item the fractal model offers an attractively simple description of the
      galaxy distribution in the observed universe, and the results
      produced by the new statistical approach in the form of the
      conditional density $\Gamma(r)$ and its derived functions are
      consistent with this fractal picture. Therefore the galaxy
      catalogues available
      so far may be considered as statistically fair samples of this
      distribution, which is contrary to the traditional lines where
      only a homogeneous sample is considered to be fair;     
\item from this new statistical approach we can see an agreement between
      various available redshift catalogues in the range of $0.1 - 10^3$
      h$^{-1}$ Mpc, as shown in figure \ref{fig-7-pmsl1996}, without any
      tendency to homogeneity at this scale. These redshift catalogues obey
      a density power law decay and have fractal dimension $D \approx 2$;
\item a three dimensional galaxy distribution, which has fractal properties
      when projected on an angular distribution, appears relatively homogeneous
      at some large scale;
\item the function $\xi(r)$ becomes an appropriate tool only in the limiting
      case where the sample size is much bigger than the new correlation
      scale, that is, when $\lambda_0 \ll R_ S$, so that the orthodox
      correlation length $r_0$ is related only to the new correlation
      length $\lambda_0$, and not to the sample size $R_S$.
\end{enumerate}

Some authors, although recognizing the problems of the standard
analysis, have argued against a fractal distribution from a somewhat 
unconvincing perspective. For example, Davis et al. (1988) have
claimed that although their analysis confirm the dependence of the
correlation length with the sample size, this effect cannot be
explained by Pietronero's single fractal model because they found $r_0$
to be approximately proportional to the square root of the sample size,
while it is linear in Pietronero's picture. Davis et al. (1988)
nevertheless have offered only statistical explanations where it is not
always clear what are the hidden hypotheses assumed by them. This point
is of especial concern due to the current widespread practice of
``correcting'' the samples in order to ``improve'' them for ``better''
statistics. The problem with this sort of practice is that the homogeneous
hypothesis is often implicit, and this fact considerably weakens that
kind of statistical explanations. \footnote{ \ See Coleman and Pietronero
(1992) for criticisms of this sort of practice and some examples of these
``corrections'' where the homogeneous hypothesis is actually implicit
in those procedures.}

Another important point that ought to be said again about the fractal
system as defined above is its total incompatibility with the two-point
correlation function $\xi(r)$. The problem lies in the fact that this
sort of fractal system does not have a well defined average density,
at least in between upper
and lower cutoffs. Ruffini, Song and Taraglio (1988) have, nonetheless,
pointed out that there are fractal systems in which the average density
does tend to a finite and non-zero value, which led them to suppose that
this should be the case in cosmology. Although this is a reasonable
point to  raise, the problem with this argument, as we see it, is
again the {\it a priori} assumption that this must be the case in
cosmology, i.e., that in cosmology the average density ought to tend to
a finite non-zero limit for very large distances. Obviously, behind this
idea lies the homogeneous hypothesis which Ruffini, Song and
Taraglio (1988) try to incorporate into the fractal cosmology, but, once
again, by means of an untested argument. From a relativistic perspective,
at first it would seem reasonable to think that this should be the case
because the standard Friedmannian cosmology is spatially homogeneous, and
as this is the most popular relativistic cosmological model, we would
have to incorporate some sort of homogenization or upper cutoff in the
model, sooner or later. However, as shown by Ribeiro (1992, 1993, 1994)
this viewpoint may be a rather misleading approach to the problem as we
can have an interpretation of the Friedmann models where they have no well
defined average density.

The conceptual advantage of Pietronero's scenario is the absence of
this sort of a priori reasoning, which means that it has the
ability of describing properly the distribution of galaxies if it really
forms a fractal structure. In other words, this scenario is able to free
ourselves from the homogeneity hypothesis, putting us in a position
to {\it test} whether or not the galaxies are actually distributed
uniformly, rather than starting assuming it as so far has been mostly
done in the literature concerning observational cosmology. If the
distribution of galaxies does tend to
homogeneity, this will be indicated in the measurements of the fractal
dimension inasmuch as in such case it will tend to the value $D=3$.
Nevertheless, despite this advantage offered by the fractal
picture, some researchers still claim that fractals bring ``nothing new''
to the galaxy clustering problem, while others, even if accepting
fractals at small scales, insist on the need for an eventual
homogeneity at an unspecified large scale. It is our point of view that
it is best to allow this issue to be decided by the observations
themselves, by the measurement of $D$, rather than be guided, or
misguided, by untested assumptions on homogeneity as has been mostly
done so far. In such case, great care must be exercised in order to avoid
introducing in the tests the very hypotheses that they are supposed to
verify. Currently there is a lot of controversy surrounding those points
of homogenization or not at larger scales and the validity of the
methods used, with claims and counter claims on the results published
succeeding each other, and no apparent sight of a consensus being even
close to being achieved. \footnote{ \ The tip of the iceberg of this
controversy can be seen in the following papers: Coleman, Pietronero
and Sanders (1988); Peebles (1989); Calzetti and Giavalisco (1991);
Coleman and Pietronero (1992); Maurogordato, Schaeffer and da
Costa (1992); Peebles (1993, 1996); Davis (1997); Pietronero (1997);
Pietronero, Montuori and Sylos Labini (1997); Coles (1998); Sylos Labini,
Montuori and Pietronero (1998). Additional references about this debate
can also be found at {\tt http://www.phys.uniroma1.it/DOCS/PIL/pil.html.} }

\vspace{5mm}
\begin{flushleft}
{\large \bf Acknowledgements}
\end{flushleft}

We are grateful to F.\ Sylos Labini for reading the original manuscript
and for helpful comments. The financial support from FAPERJ and CNPq is,
respectively, acknowledged by MBR and AYM.

\vspace{1cm}

{\protect\large \bf References}
\begin{description}
  \item Barnsley, M.\ 1988, {\it Fractals Everywhere} (London:
        Academic Press).
  \item Baryshev, Y., Sylos Labini, F., Montuori, M.\ and Pietronero,
        L.\ 1994, {\it Vistas in Astron.}, 38, 419.
  \item B\'elair, J.\ and Dubuc, S.\ (eds.) 1991, {\it Fractal Geometry
        and Analysis} (Dordrecht: Kluwer). 
  \item Calzetti, D.\ et al. 1987, {\it Ap.\ Space Sci.}, 137, 101.
  \item Calzetti, D.\ and Giavalisco, M.\ 1991, in {\it Applying Fractals
        in Astronomy}, p.\ 119, ed.\ A.\ Heck and J.\ M.\ Perdang
	(Berlin: Springer-Verlag).
  \item Coleman, P.\ H.\ and Pietronero, L.\ 1992, {\it Phys.\ Reports},
        213, 311.
  \item Coleman, P.\ H., Pietronero, L.\ and Sanders, R.\ H.\ 1988, {\it
	A \& A}, 200, L32. 
  \item Coles, P.\ 1998, {\it Nature}, 391, 120.
  \item Davis, M.\ 1997, {\it Is the Universe Homogeneous on Large
        Scales?}, in {\it Critical Dialogues in Cosmology}, p.\ 13,
	ed.\ N.\ Turok (Singapore: World Scientific).
  \item Davis, M.\ et al. 1988, {\it ApJ}, 333, L9.
  \item de Vaucouleurs, G.\  1970, {\it Science}, 167, 1203. 
  \item Einasto, J.\, Klypin, A.\ A.\ and Saar, E.\ 1986, {\it M.\ N.\ R.\
        A.\ S.\ }, 219, 457.
  \item Falconer, K.\ 1990, {\it Fractal Geometry: Mathematical
        Foundations and Applications} (Chichester: John Wiley and Sons).
  \item Feder, J.\ 1988, {\it Fractals} (New York: Plenum Press).
  \item Geller, M.\ 1989, {\it The Large-Scale Distribution of Galaxies},
	in {\it Astronomy, Cosmology and Fundamental Physics}, p.\ 83,
	eds.\ M.\ Caffo, R.\ Fanti, G.\ Giacomelli and A.\ Renzini
	(Dordrecht: Kluwer Academic Publishers).
  \item de Lapparent, V., Geller, M.\ J.\ and Huchra, J.\ P.\  1986,
        {\it ApJ}, 302, L1. 
  \item Mandelbrot, B.\ B.\ 1983, {\it The Fractal Geometry of Nature}
	(New York: Freeman).
  \item Maurogordato, S., Schaeffer, R.\ and da Costa, L.\ N.\ 1992, {\it
        ApJ}, 390, 17. 
  \item Peebles, P.\ J.\ E.\ 1980, {\it The Large-Scale Structure of the
        Universe} (Princeton University Press).
  \item Peebles, P.\ J.\ E.\ 1989, {\it Physica}, 38D, 273, [in 
        {\it Fractals in Physics, Proc.\ of the International Conference 
	honouring Benoit B.\ Mandelbrot on his 65th birthday}, p.\ 273,
	eds.\ A.\ Ahorony and J.\ Feder (North Holland)].  
  \item Peebles, P.\ J.\ E.\ 1993, {\it Principles of Physical Cosmology}
        (Princeton University Press).
  \item Peebles, P.\ J.\ E.\ 1996, {\it circular letter to L.\ Pietronero}:
        available
        at \newline {\tt http://www.phys.uniroma1.it/DOCS/PIL/peebles.html}.
  \item Peitgen, H-O, J\"{u}rgens, H.\ and Saupe, D.\ 1992, {\it Fractals
        for the Classroom - Part One: Introduction to Fractals and
	Chaos} (New York: Springer-Verlag).
  \item Pietronero, L.\ 1987, {\it Physica}, 144A, 257. 
  \item Pietronero, L.\ 1988, {\it Fractals in Physics: Introductory
        Concepts}, in {\it Order and Chaos in Nonlinear Physical
	Systems}, p.\ 277, ed.\ S.\ Lundqvist, N.\ H.\ March and M.\ P.\
	Tosi (New York: Plenum Press).
  \item Pietronero, L.\ 1997, {\it circular letter to P.\ J.\ E.\ Peebles}:
        available
        at \newline {\tt http://www.phys.uniroma1.it/DOCS/PIL/lett-lp.ps}.
  \item Pietronero, L., Montuori, M.\ and Sylos Labini, F.\ 1997, {\it
        On the Fractal Structure of the Visible Universe}, in {\it Critical
	Dialogues in Cosmology}, p.\ 24, ed.\ N.\ Turok (Singapore:
	World Scientific).
  \item Raine, D.\ J.\ 1981, {\it The Isotropic Universe: An Introduction
        to Cosmology}, (Bristol: Adam Hilger).
  \item Ribeiro, M.\ B.\ 1992, {\it ApJ}, 395, 29.
  \item Ribeiro, M.\ B.\ 1993, {\it ApJ}, 415, 469.
  \item Ribeiro, M.\ B.\ 1994, {\it Relativistic Fractal Cosmologies}, in {\it 
        Deterministic Chaos in General Relativity}, p.\ 269, ed.\ D.\ Hobill
	{\it et al.} (New York: Plenum Press).
  \item Ribeiro, M.\ B.\ 1995, {\it ApJ}, 441, 477.
  \item Ruffini, R., Song, D.\ J.\ and Taraglio, S.\ 1988, {\it A \& A},
        190, 1.
  \item Saunders, W.\ et al. 1991, {\it Nature}, 349, 32. 
  \item Sylos Labini, F., Gabrielli, A., Montuori, M.\ and Pietronero,
        L.\ 1996, {\it Physica A}, 226, 195.
  \item Sylos Labini, F., Montuori, M.\ and Pietronero, L.\ 1998,
        {\it Phys.\ Reports}, 293, 61.   
  \item Takayasu, H.\ 1990, {\it Fractals in the Physical Sciences}
        (Manchester University Press).
  \item Wertz, J.\ R.\ 1970, {\it Newtonian Hierarchical Cosmology}, 
        PhD thesis (University of Texas at Austin). 
  \item Wertz, J.\ R.\ 1971, {\it ApJ}, 164, 227. 
\end{description}

\end{document}